\documentclass[10pt,conference]{IEEEtran}
\IEEEoverridecommandlockouts
\usepackage[font=small,labelfont=bf,tableposition=top]{caption}
\usepackage{hyperref}
\usepackage{boldline}
\usepackage{color,colortbl}
\usepackage{bigstrut}
\usepackage[ruled]{algorithm2e}
\usepackage{amsmath,amsfonts}
\usepackage{amsmath}
\usepackage{array}
\usepackage{algorithmic}
\usepackage{balance}
\usepackage{blindtext}
\usepackage{booktabs}
\usepackage{caption}
\usepackage{hyperref}
\usepackage[noabbrev]{cleveref}
\usepackage{cite}
\usepackage{comment}
\usepackage{enumitem}
\usepackage{eqparbox}
\usepackage{fancybox}
\usepackage{fancyvrb}
\usepackage{framed}
\usepackage{graphicx}
\usepackage{ifthen}
\usepackage{listings}
\usepackage{mathrsfs}
\usepackage{mdwmath}
\usepackage{mdwtab}
\usepackage{multirow}
\usepackage{pifont}
\usepackage{stfloats}
\usepackage{textcomp}
\usepackage{times}
\usepackage{url}
\usepackage{xspace}
\usepackage[normalem]{ulem}
\usepackage[framemethod=TikZ]{mdframed}
\usepackage{caption}
\usepackage{subcaption}
\usepackage{tabularx}
\usepackage[switch]{lineno}

\usepackage{titlesec}

\usepackage{tcolorbox}
\tcbuselibrary{listings,skins}
 
\usepackage{mathtools}
\usepackage{blindtext}
\usepackage{lstautogobble}

\DeclareGraphicsExtensions{.pdf,.jpeg,.png}
\graphicspath{{images/}}

\titlespacing\section{0pt}{8pt plus 4pt minus 2pt}{4pt plus 2pt minus 2pt}
\newcommand{\rom}[1]{\uppercase\expandafter{\romannumeral #1\relax}}

\newcommand{\eg}{\hbox{\emph{e.g.,}}\xspace}
\newcommand{\ie}{\hbox{\emph{i.e.,}}\xspace}

\definecolor{gray50}{gray}{.5}
\definecolor{gray40}{gray}{.6}
\definecolor{gray30}{gray}{.7}
\definecolor{gray20}{gray}{.8}
\definecolor{gray10}{gray}{.9}
\definecolor{gray05}{gray}{.95}
\definecolor{gray01}{gray}{.97}

\newlength\Linewidth
\def\findlength{\setlength\Linewidth\linewidth
\addtolength\Linewidth{-4\fboxrule}
\addtolength\Linewidth{-3\fboxsep}
}
\newenvironment{examplebox}{\par\begingroup
  \setlength{\fboxsep}{3pt}\findlength
  \setbox0=\vbox\bgroup\noindent
  \hsize=0.90\linewidth
  \begin{minipage}{0.90\linewidth}\normalsize}
    {\end{minipage}\egroup
    \textcolor{gray20}{\fboxsep1.2pt\fbox
     {\fboxsep5pt\colorbox{gray05}{\normalcolor\box0}}}
    \endgroup\par\noindent
    \normalcolor\ignorespacesafterend}

\usetikzlibrary{shadows}
\usepackage{graphics}
\newmdenv[
    tikzsetting= {fill=blueish},
    skipabove=0.33em,
    skipbelow=0.33em,
    linewidth=1pt,
    innerleftmargin=4pt,
    innerrightmargin=4pt,
    innertopmargin=2pt,
    innerbottommargin=2pt,
    linecolor=gray95,
    roundcorner=2pt, 
    shadow=true,
    shadowsize=4pt,
    shadowcolor=gray95
]{questionbox}

\newmdenv[
    tikzsetting= {fill=greenish},
    skipabove=0.33em,
    skipbelow=0.33em,
    linewidth=1pt,
    innerleftmargin=4pt,
    innerrightmargin=4pt,
    innertopmargin=2pt,
    innerbottommargin=2pt,
    linecolor=gray95,
    roundcorner=2pt, 
    shadow=true,
    shadowsize=4pt,
    shadowcolor=gray95
]{answerbox}

\newmdenv[
    skipabove=0.33em,
    skipbelow=0.33em,
    innerleftmargin=4pt,
    innerrightmargin=4pt,
    innertopmargin=2pt,
    innerbottommargin=2pt,
]{lessonbox}

\usepackage{tikz}

\newenvironment{lesson}
{
    \begin{lessonbox}
    Lessons Learned:\\
}
{\end{lessonbox}}

\newenvironment{result}
{\begin{answerbox}}
{\end{answerbox}}

\newenvironment{question}
{\begin{questionbox}}
{\end{questionbox}}

\definecolor{javared}{rgb}{0.6,0,0} % for strings
\definecolor{javagreen}{rgb}{0.25,0.5,0.35} % comments
\definecolor{javapurple}{rgb}{0.5,0,0.35} % keywords
\definecolor{javadocblue}{rgb}{0.25,0.35,0.75} % javadoc

\lstdefinestyle{basejava}{
  language=java,
  showstringspaces=false,
  basicstyle=\scriptsize\ttfamily,
  keywordstyle=\bfseries\color{javapurple},
  commentstyle=\itshape\blue,
  identifierstyle=\blue,
  frame=none,
  backgroundcolor=\color{white},
}

\lstdefinestyle{CustomJava}{
  belowcaptionskip=\baselineskip,
  breaklines=true,
  xleftmargin=\parindent,
  language=java,
  showstringspaces=false,
  basicstyle=\scriptsize\ttfamily,
  keywordstyle=\bfseries\color{javapurple},
  commentstyle=\itshape\blue,
  identifierstyle=\blue,
  belowskip=1pt,
  frame=shadowbox,
  backgroundcolor=\color{gray01},
  gobble=0
}

\lstdefinestyle{codit}{
  belowcaptionskip=\baselineskip,
  breaklines=true,
  language=java,
  showstringspaces=false,
  basicstyle=\scriptsize\ttfamily,
  keywordstyle=\bfseries\color{javapurple},
  commentstyle=\itshape\blue,
  identifierstyle=\blue,
}

\newtcblisting{mylisting}[2][]{
    arc=3mm,
    listing only, 
    listing style=codit,
    title=#2,
    #1
    }

\newcommand\blue[1]{\textcolor[rgb]{0.00,0.00,1.00}{{#1}}}

\definecolor{blueish}{RGB}{250, 250, 255}
\definecolor{greenish}{RGB}{250, 255, 250}
\definecolor{redish}{RGB}{255, 200, 200}

\definecolor{gray05}{gray}{0.95}
\definecolor{gray08}{gray}{0.92}
\definecolor{gray10}{gray}{0.90}
\definecolor{gray12}{gray}{0.88}
\definecolor{gray15}{gray}{0.85}
\definecolor{gray18}{gray}{0.82}
\definecolor{gray20}{gray}{0.80}
\definecolor{gray25}{gray}{0.75}
\definecolor{gray30}{gray}{0.70}
\definecolor{gray35}{gray}{0.65}
\definecolor{gray40}{gray}{0.60}
\definecolor{gray45}{gray}{0.55}
\definecolor{gray50}{gray}{0.50}
\definecolor{gray55}{gray}{0.45}
\definecolor{gray60}{gray}{0.40}
\definecolor{gray65}{gray}{0.35}
\definecolor{gray70}{gray}{0.30}
\definecolor{gray75}{gray}{0.25}
\definecolor{gray80}{gray}{0.20}
\definecolor{gray85}{gray}{0.15}
\definecolor{gray90}{gray}{0.10}
\definecolor{gray95}{gray}{0.05}
\xspace%

\newcommand{\Comment}[1]{}

\newcommand{\citep}[1]{\cite{#1}}

\newtcbox{\inlinebox}[1][]{enhanced,
 box align=base,
 nobeforeafter,
 colback=blueish,
 size=small,
 left=0pt,
 right=0pt,
 boxsep=2pt,
 #1}

\newcommand{\RQ}[2]{%
    {
    \small
    \begin{question}
        \label{rq-#1}
        \noindent\textbf{{RQ{#1}.~#2}}
    \end{question}}
}

\newcommand{\RS}[2]{%
    {\small
    \begin{result}
        \textbf{\hyperref[rq-#1]{Result {#1}}:~}{\emph {#2}}%
    \end{result}}
}

\renewcommand{\cref}[1]{\Cref{#1}}

\newcommand{\tool}{\textsc{Reinforest}\xspace}

\newcommand{\rqa}{How does \tool's performance compare to the performance of other cross-language code search techniques?}
\newcommand{\rqb}{Does \tool's methodology and performance generalize across different models?}

\newcommand{\rqc}{Does including semantic similarity scores during training improve code search?}

\newcommand{\rqd}{How does changing the number of positive and negative comparison samples available for training effect \tool's performance?}

\usepackage[turnoff]{notes}
\IEEEtriggeratref{46}
\begin{document}

\title{\tool : Reinforcing Semantic Code Similarity for Cross-Lingual Code Search Models}

% Author macro is renewed to empty if notes is turned off. Make sure to turn off the notes before submission.
\author{
    \IEEEauthorblockN{Anthony Saieva$^\dagger$ \thanks{$^\dagger$Equal Contribution}}
    \IEEEauthorblockA{
        {Columbia University}\\
        New York, NY, USA \\
        \href{saikatc@cs.columbia.edu}{ant@cs.columbia.edu}
    }
    \and
    \IEEEauthorblockN{Saikat Chakraborty$^\dagger$}
    \IEEEauthorblockA{
        {Microsoft Research}\\
        Redmond, WA, USA \\
        \href{saikatc@microsoft.com}{saikatc@microsoft.com}
    }
    \and
    \IEEEauthorblockN{Gail Kaiser}
    \IEEEauthorblockA{
        {Columbia University}\\
        New York, NY, USA \\
        \href{kaiser@cs.columbia.edu}{kaiser@cs.columbia.edu}
    }
}
\maketitle
\begin{abstract}
This paper introduces a novel code-to-code search technique that enhances the performance of Large Language Models (LLMs) by including both static and dynamic features as well as utilizing both similar and dissimilar examples during training. We present the first-ever code search method that encodes dynamic runtime information during training without the need to execute either the corpus under search or the search query at inference time and the first code search technique that trains on both positive and negative reference samples. To validate the efficacy of our approach, we perform a set of studies demonstrating the capability of enhanced LLMs to perform cross-language code-to-code search.

Our evaluation demonstrates that the effectiveness of our approach is consistent across various model architectures and programming languages. We outperform the state-of-the-art cross-language search tool by up to 44.7\%. Moreover, our ablation studies reveal that even a single positive and negative reference sample in the training process results in substantial performance improvements demonstrating both similar and dissimilar references are important parts of code search. Importantly, we show that enhanced well-crafted, fine-tuned models consistently outperform enhanced larger modern LLMs without fine tuning, even when enhancing the largest available LLMs highlighting the importance for open-sourced models.

 To ensure the reproducibility and extensibility of our research, we present an open-sourced implementation of our tool and training procedures called \tool.

\end{abstract}

\begin{IEEEkeywords}
Large Language Models, Semantic Clones, Semantic Code Search
\end{IEEEkeywords}

\section{Introduction}
\label{sec:intro}

With the rise of powerful large language models (LLM's) since the original BERT \cite{DBLP:conf/naacl/DevlinCLT19} paper was published, machine learning applied to source code analysis tasks has become increasingly popular in the last few years.  At the same time as large public repositories of code have become available \cite{vasilescu2019state}, the potential of code-to-code search has gained new significance to aid in common software maintenance and development tasks like code migration \cite{kim2017mining}, transpilation \cite{zhou2018automatic}, code repair \cite{long2018cross}, bug detection \cite{kim2013automatic}, education \cite{yao2019exploring}, and refactoring \cite{8530022}.    

\begin{figure}[htbp]
  \centering
  \begin{subfigure}[b]{0.48\textwidth}
    \includegraphics[width=\textwidth]{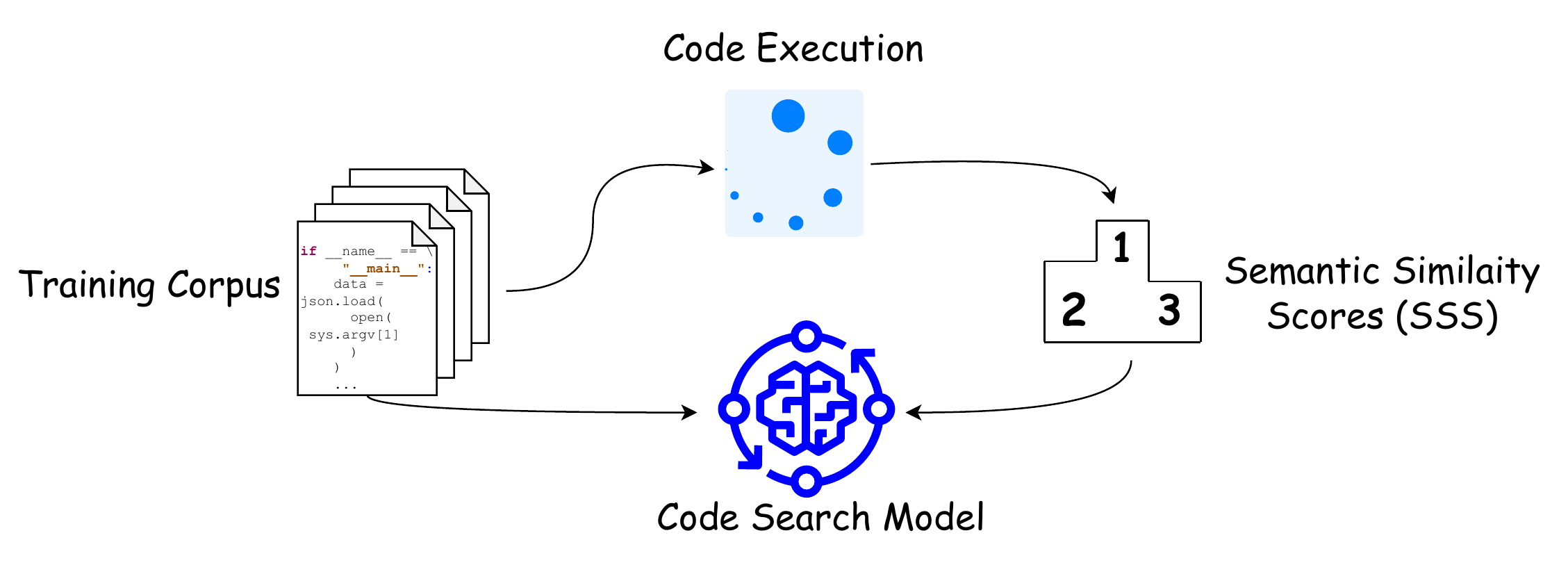}
    \caption{Overview of the training process}
    \label{fig:training}
  \end{subfigure}
  \hfill
  \begin{subfigure}[b]{0.48\textwidth}
    \includegraphics[width=0.9\textwidth]{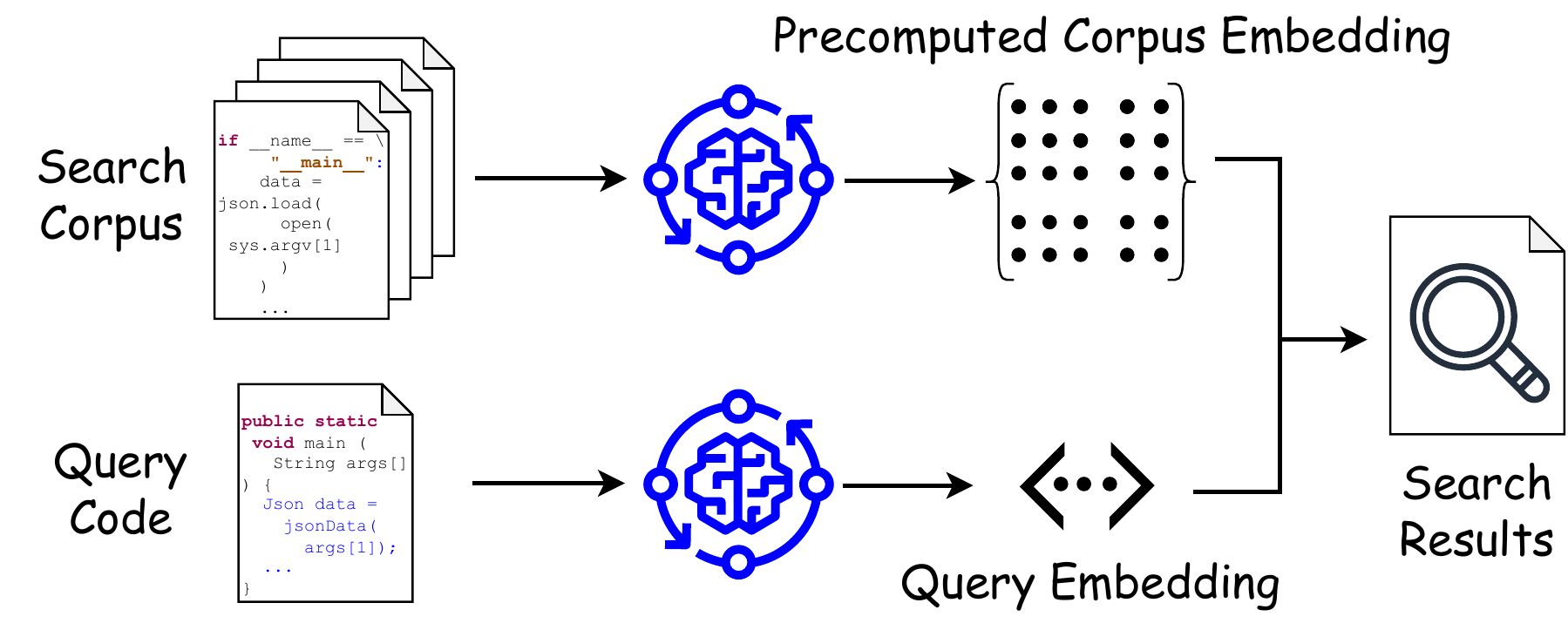}
    \caption{Overview of the inference process}
    \label{fig:inference}
  \end{subfigure}
  \caption{Overview of the training and inference processes}
  \label{fig:overview}
\end{figure}

In this paper, we present a series of novel techniques that enhance LLMs for code-to-code search and related tasks. We introduce a new code-to-code search technique that encodes both source code and dynamic runtime information during training without needing to execute anything for inference from either the search corpus or the search query. Figure \ref{fig:overview} shows the general premise. First, during training the model learns both static and dynamic similarity from the training corpus, maximizing the distance between dissimilar code and minimizing the distance between similar code. We generate a single dynamic similarity feature we call the Semantic Similarity Score (SSS) before training that involves executing as much of the training corpus as possible with the same inputs and comparing the outputs of functions that have the same input types. This is the only time code is executed. Then to put this model to use, the model is used to create a precomputed embedding of the search corpus, and finally at inference time the model embeds the search query so a simple comparison to the precomputed embeddings yields the search results. Traditionally dynamic analysis techniques require executing code during training as well as executing the search corpus and search query during the search procedure. This leads to problems scaling and all of the environment-related issues that come with running code in practice \cite{yang2021practical}. In contrast our approach only incurs the overhead of executing training set code in a controlled environment, and once the model is distributed code never needs to be executed again. Notably, our approach still accommodates circumstances when parts or all of the training set are not executable. 

To validate the effectiveness of our approach, we considered the problem of cross-language search in which the query and corpus are from different languages, as in COSAL \cite{mathew2021cosal}. We felt this would best demonstrate the effectiveness of our technique since we are most concerned with behavioral similarity and searching across syntactically different languages means the model cannot rely on source code similarity.

We conduct a series of ablation studies that demonstrate the ability of LLMs to infer dynamic behavior under a variety of circumstances even when inference is based solely on static information. These circumstances include multiple query and corpus languages, various underlying LLM models both when the models are open-sourced and proprietary, when dynamic runtime information is available and without runtime information, and finally varying the number of both positive and negative reference samples in training. Our evaluation shows that our technique performs considerably better than state-of-the-art cross-language search techniques by up to 44.7\% on the benchmark Atcoder \cite{nakamura2021atcoder} dataset, which consists of Java and Python solutions corresponding to 361 problems from 83 programming contests, and maintains near or better performance with a variety of LLM architectures including CodeBERT\cite{feng2020codebert} and OpenAI's codex models \cite{brown2021language}. even when the LLM's cannot be fine-tuned. It is worth noting that we find that a smaller fine-tuned LLM outperforms significantly larger proprietary LLM's highlighting the value and importance of open-sourced models for the research community. Further, we find that as long as there is at least one positive reference and one negative reference available during training our technique performs effectively compared to benchmarks, allowing for sparser datasets as what might be found in practice \cite{ray2016onward} and demonstrating the importance of considering both similar and dissimilar reference samples.

We make the following contributions:
\begin{enumerate}
    \item The first code-to-code search technique that encodes dynamic runtime information during training without the need to execute any code from either the search corpus or the search query during inference.
    \item The first code-to-code search technique that considers both similar and dissimilar reference samples during training.
    \item A set of ablation studies that demonstrate the ability of LLM's to infer dynamic behavior during cross-language search even though inference is only based on static information.
    \item An evaluation that shows our technique remains effective across many model architectures.
    \item Ablation studies showing even a single positive and single negative reference sample in training results in effective performance.
    \item An evaluation showing well-crafted fine-tuned models out-perform larger non fine-tuned LLM models even with the largest LLM's.
    \item An open-sourced implementation of our tool and training procedures called \tool~\footnote{\url{https://github.com/reinforest-team/REINFOREST}} that represents a new state-of-the-art in cross-language code search.
\end{enumerate}
%
% \cite{chakraborty2018tree2tree}
\section{Background}
\label{sec:background}

\subsection{Code Search}

Code to code search can be defined broadly as taking a code sample (the query), and searching through a corpus of other available code samples (the search corpus), and finding code clones in the search corpus matching the query. Cross-language code to code search deals specifically with situations when the query and corpus are written in different languages. Consider a situation in which Jane the developer is tasked with writing code in a language she is relatively new to or a language that lacks supporting documentation. Rather than writing without a guide, with cross-language code to code search Jane can make a query with a reference sample from a language she knows well and look for similar samples in a corpus of the language she is not as comfortable with.  

Su et al. \cite{7503720} define four type of code clones:
\begin{itemize}
    \item \textbf{Type 1}: Identical code fragments, except for variations in whitespace, comments, or formatting.
    \item \textbf{Type 2}: Syntactically identical code fragments, with differences in variable, constant, or function names.
    \item \textbf{Type 3}: Similar code fragments with modified statements, function calls, or control structures.
    \item \textbf{Type 4}: Semantically identical code fragments with the same functionality but different syntactic structures.    
\end{itemize}

Traditionally, type four clones are the most difficult to detect as the source code may offer little if any useful context, and that is the type we are most concerned with in this work. If our technique successfully learns code behavior we should be able to perform code search even when the syntactic structures are from different languages.

Code clones arise from several factors, including copy-pasting, code reuse, and adherence to coding patterns, and code search techniques are essential tools for developers, as they help locate and reuse code snippets in large codebases. Existing static code search techniques can broadly be classified into three categories 1) text-based approaches: These methods rely on traditional information retrieval techniques such as keyword search or string matching to find relevant code snippets, 2) structural approaches: These methods leverage the syntactic and semantic structure of source code, such as abstract syntax trees, to find matching or similar code fragments, and 3) machine learning-based approaches, which leverage both text and semantic structural encoding along with machine learning algorithms to find clones. Traditionally code search has been limited to query and corpus of code from the same language, but in recent years this has expanded to search code bases where the query and corpus may be of different languages \cite{kamei2013cross, wu2016cross, xia2018neural} as code search algorithms get better at distilling semantic information from the source code text.

Additionally techniques that include information from dynamic analysis like execution traces \cite{10.1145/2950290.2950321} or input output pairs \cite{7503720} leverage additional context to find code clones that were undetected by static analysis. This is especially important in type four clones, which only concerns behavioral similarity. 

On the one hand static code search techniques do not require code execution, making them highly scalable for searching large codebases processing vast amounts of source code quickly, and providing fast search results to developers \cite{rastogi2015prose}. They are also easy to implement as text-based and structural approaches in static code search are generally simpler to implement than dynamic methods, and require no execution environment nor runtime setup making them more portable and applicable to a wide range of software projects. These practical benefits come at the cost of less context than what is available to dynamic techniques, which results in lower performance. On the other hand, dynamic techniques offer improved accuracy and context-awareness but can be more complex and less scalable \cite{marcum2010comparative}.

By including dynamic information during training where possible, and not requiring it at  inference time, we aim to bridge the gap between dynamic and static code search bringing together both the convenience of static methods and the enhanced context of dynamic methods. 

% \saikat{background on encoder llms}

\subsection{Encoder Models for source code understanding}

With the advancement of Deep Learning models and techniques, models have been trained to "understand" source code. Given a code snippet ${c}$ and an encoder $Enc$, the model generates a $d$ dimensional vector representation of the code ${Enc}(c) \in \mathbb{R}^d$. 
This vector represents the lexicographic, syntactic, and semantic information in the code. Such vector embedding can be subsequently used for various different tasks including, but not limited to, classification, search, ranking, etc. The encoder can be an open-source model, such as CodeBERT~\cite{feng2020codebert} or a closed-source embedding API such as OpenAI's codex embedding~\cite{openai_emb}. For a highly effective pretrained model, the embeddings themselves should be sufficient for tasks related to code understanding however, as previous research has shown~\cite{nitin2021direct}, the embeddings need to be tuned depending on the downstream tasks. In the case of open-source models for embedding, where we can access the models' weights, the models can be fine-tuned as part of the downstream tasks. For embeddings coming from closed-source models, further models can be built on top of the initial embeddings, but the embeddings cannot be changed. 

\section{\tool}
\label{sec:tool}

In this section, we describe the code search technique we propose in \tool. We describe \tool as two disjoint steps -- (a) Training and (b) Online query embedding and search. At the code level, \tool consists of two different encoders -- query code encoder ($E_q$) and the document code encoder ($E_d$). Each of these, given an input code $x$, generates a fixed-sized vector representation of $x$.

\begin{figure*}[t]
    \centering
    \includegraphics[width=0.85\textwidth]{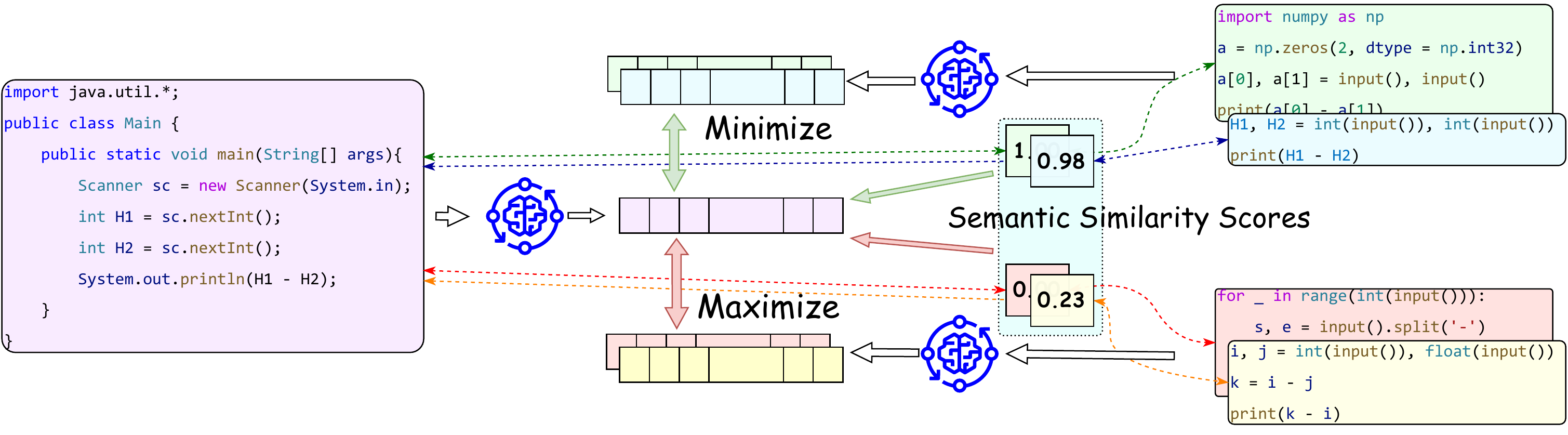}
    \caption{Training process in \tool}
    \label{fig:overview_training}
\end{figure*}

\subsection{Training \tool}

We train \tool aiming to optimize for two different orthogonal goals. First, we would want \tool to transform embeddings generated by the query code encoder ($E_q$) and the document code encoder ($E_d$) in such a way that minimizes the distance between related code and maximizes the distance between unrelated code. Our second goal is to endow \tool with dynamic code similarity information during training so that, during inference (search), we do not have to calculate the semantic (I/O) similarity between the query code and all other code in the database. \Cref{fig:overview_training} shows an overview of \tool's training process.

{\bf \textit{Contrastive Training}.} To achieve the first goal, we take inspiration from contrastive learning~\cite{chen2020simple}. We assume the presence of a dataset $\mathcal{D}$ consisting of a set of tuples $(c_s, \mathcal{C}_p, \mathcal{C}_n)$. $c_s$ is a piece of code written in the source language $s$ (\eg Java).  $\mathcal{C}_p$ and $\mathcal{C}_n)$ are the set of code written in the target language (\eg Python). The set $\mathcal{C}_p = \{p_{t1}, p_{t2}, ..., p_{tk}\}$ are the solutions to the same problem as $c_s$. In contrast, $\mathcal{C}_n = \{n_{t1}, n_{t2}, ..., n_{tk}\}$ are solutions to a different problem than $c_s$. We refer to the code samples $\mathcal{C}_p$ as positive samples and code samples $\mathcal{C}_n$ as negative samples hereafter. Intuitively, we would want to bring the embeddings of each element in $\mathcal{C}_p$ closer to the embedding of $c_s$, and we would want to push the embedding of $\mathcal{C}_n$'s elements far away from $c_s$'s embedding. 

First, we encode $c_s$ with the query code encoder, $E_q$, and both the $\mathcal{C}_p$ and $\mathcal{C}_n$ sets with the document code encoder $E_d$. Formally, $R_s = E_q(c_s)$, $\mathcal{R}_p = \{E_d(p_{ti})~|~p_{ti} \in \mathcal{C}_p\}$, and $\mathcal{R}_n = \{E_d(n_{ti})~|~n_{ti} \in \mathcal{C}_n\}$, where $R_s$ is the embedding of the query code, $\mathcal{R}_p$, and $\mathcal{R}_n$ are the sets of embedding corresponding to positive and negative samples, respectively. Then we compute the cosine similarity between the query code and each of the positive and negative samples. We compute the cosine similarity between two vectors, $a$ and $b$ as, 
\begin{equation}
    sim(a, b) = \frac{|a\cdot b|}{|a|*|b|}
    \label{similarity}
\end{equation}

Our next goal is to simultaneously maximize the similarity between $c_s$ and all elements in $\mathcal{C}_p$ and minimize the similarity between $c_s$ and all elements in $\mathcal{C}_n$. We achieve such an objective by minimizing the following loss function,
\begin{multline}
    \mathcal{L} = \sum_{(c_s, \mathcal{C}_p, \mathcal{C}_n) \in \mathcal{D}}{\left(\sum_{r^p_{ti} \in \mathcal{R}_p}{\left(l_p - sim\left(R_s, r^p_{ti}\right)\right)^2}\right)} + \\
    {\left(\sum_{r^n_{ti} \in \mathcal{R}_n}{\left(l_n - sim\left(R_s, r^n_{ti}\right)\right)^2}\right)} 
    \label{eqn:loss_function}
\end{multline}
Here, $l_p$ and $l_n$ are target labels for positive and negative samples. The values of $l_p$ and $l_n$ should depend on the similarity function being used in the system. For an unbounded similarity function, $l_p$ and $l_n$ should be set to $\infty$ and $-\infty$, respectively. In the case of \tool, we set $l_p = 1$ and $l_n = 0$, since these values are the maxima and minima of the similarity function we use (\cref{similarity}).

{\bf \textit{Semantic Similarity Score}} As our second training goal is to endow the models with the knowledge about semantic (I/O) similarity, we assume a function $\mathcal{S}_{io}(c_s, c_t)$, which computes the semantic similarity between $c_s$ and $c_t$. We endow the model with such semantic similarity by modifying the loss function in \Cref{eqn:loss_function} as 
\begin{multline}
    \mathcal{L'} = \sum_{(c_s, \mathcal{C}_p, \mathcal{C}_n) \in \mathcal{D}}{\left(\sum_{p_{ti}, r^p_{ti}}{\left(\mathbf{l'_p(c_s, p_{ti})} - sim\left(R_s, r^p_{ti}\right)\right)^2}\right)} + \\
     {\left(\sum_{n_{ti}, r^n_{ti}}{\left(\mathbf{l'_n(c_s, n_{ti})} - sim\left(R_s, r^n_{ti}\right)\right)^2}\right)}
    \label{eqn:loss_function_modified}
\end{multline}
\begin{equation}
    \mathbf{l'_p(c_s, p_{ti})} = (1 - \alpha) l_p + \alpha \mathcal{S}_{io}(c_s, p_{ti})
\end{equation}
\begin{equation}
    \mathbf{l'_n(c_s, n_{ti})} = (1 - \alpha) l_n + \alpha \mathcal{S}_{io}(c_s, n_{ti})
\end{equation}
Here, $\alpha$ represents the relative importance of semantic training. In an ideal scenario, where we could compute the semantic similarity between all pairs of $c_s$ and $c_t$, we argue that setting $\alpha$ to 1.0 would result in the best performance. However, it is practically infeasible to assume that for all such pairs of $c_s$ and $c_t$, $\mathcal{S}_{io}(c_s, c_t)$ exists. Thus, we empirically found best validation performance by setting $\alpha$ to $0.2$. Regardless, $\alpha$ remains a user-defined hyperparameter in \tool's implementation. 

\begin{figure}[t]
     \centering
     \begin{subfigure}[b]{\linewidth}
         \centering  \includegraphics[width=.9\textwidth]{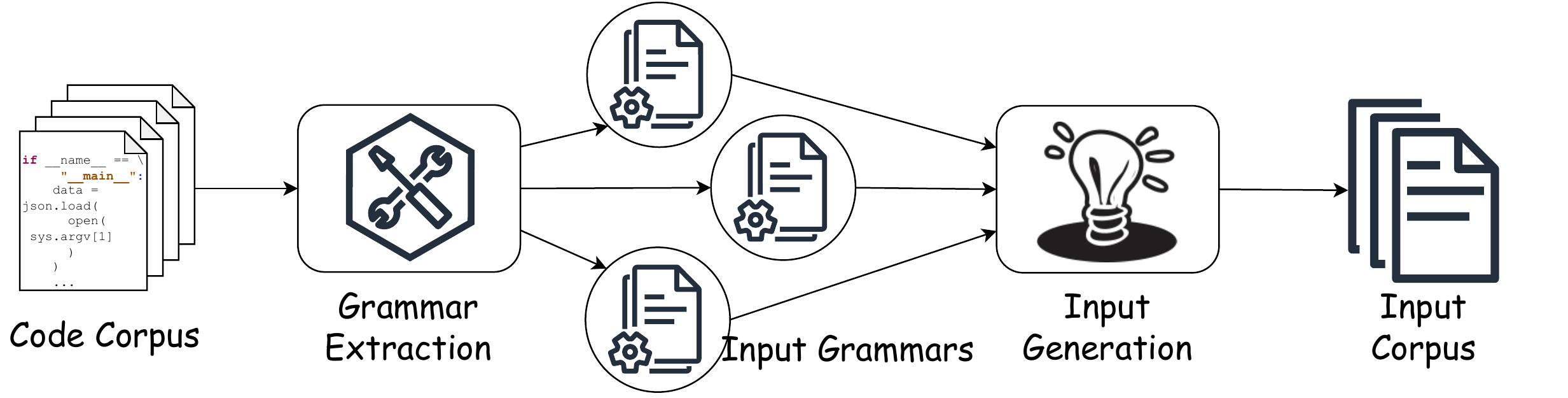}
         \caption{Corpus generation}
         \label{fig:corpus_generation}
     \end{subfigure}

    \vspace{5mm}
     \begin{subfigure}[b]{\linewidth}
         \centering
         \includegraphics[width=.9\textwidth]{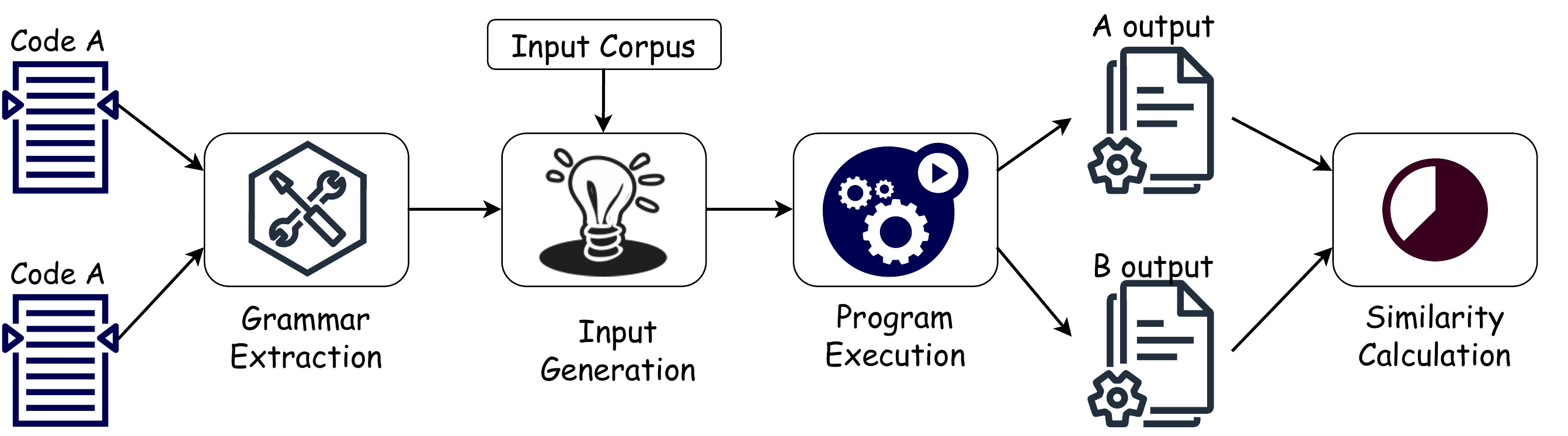}
         \caption{Similarity Calculation}
         \label{fig:similarity_calculation}
     \end{subfigure}
\caption{Similarity Score Processes}
\label{fig:similarity_figures}
\end{figure}

{\bf \textit{Semantic Similarity Score Calculation For Training}.} \tool's architecture and process uses a semantic similarity score calculation based on runtime information where possible during training. We use SLACC's\cite{9283951} similarity scoring methodology,  which is simply the number of matching outputs divided by the number of inputs. Figure \ref{fig:similarity_figures} outlines our two process procedures for generating similarity scores. First, as shown in Figure \ref{fig:corpus_generation}, we generate the input corpus from the code corpus. For any piece of code, we statically analyze the system calls and derive the structure of the input for that particular piece of code. After extracting all of the input structures in the corpus, we generate a large number of inputs of random values for each structure based on primitive types and create an input corpus. Then as shown in \ref{fig:similarity_calculation} for any two samples, we extract the input structures, run them with the inputs that match their input structure, and then perform the similarity calculation described earlier. As in SLACC, if two pieces of code have different input structures, they receive a similarity score of zero. 
% In the event that either piece of code has an error at runtime, then they receive a score of ERR (we used -1). This line is not necessary, since I convert those -1 to 0. 

\subsection{Code Search With \tool}

Once the training process finishes, we have two trained encoders, $E_q$ and $E_d$, to encode queries and documents, respectively. Immediately after the training, we encode all the code in the search database with $E_d$. Since the encoded representations are real-valued vectors, indexing tools such as FAISS~\cite{johnson2019billion} can store them in the indexed database for efficient search.  For every query code $c_s$, we generate the query embedding $R_s$ with the query encoder $E_s$. Then we search in the embedding database based on the similarity score in \cref{similarity} and return the top $n$ candidate with the highest similarity scores.

\section{Experimental Design}
\label{sec:method}

\subsection{Research Questions}
To evaluate our technique we implemented it in a tool called \tool and we considered four main questions. The most important and broadest of which is simply:
\RQ{1}{\rqa}

% reimplemented cosal 
% ast -- reimplemented cosal generic tree structure
% 

Still better performance could mean that we simply rely on the power of modern LLM's. If, however, our tool continues to show state-of-the-art performance with multiple LLM's, and our additional training method always improves performance this demonstrates that our technique is in fact the source of any performance advantages shown in RQ1. So we ask:  
\RQ{2}{\rqb}

Even if RQ2 shows performance improvements across multiple architectures, it may only be from source code analysis and the dynamic runtime information may not make a difference, so we ask:
\RQ{3}{\rqc}

Lastly, it remains unclear what impact including both positive and negative samples during training has, so we ask:
\RQ{4}{\rqd}

\subsection{Dataset}
\begin{table}[t]
    \centering
    \caption{Statistics of the dataset. The ``Average" denotes the average number of files per problem.}
    \label{tab:data_stat}
    \begin{tabular}{ll|r|r|r}
        \hlineB{2}
        & & \textbf{Train} & \textbf{Valid} & \textbf{Test} \bigstrut\\
        \hline
         \multicolumn{2}{c|}{Number of Problems} & 287 & 37 & 37\bigstrut\\
        \hline
         \multirow{2}{*}{Number of Java files} & Total & 14413 & 2533 & 1698 \bigstrut\\
          & Average & 50.22 & 68.50 & 48.89\bigstrut\\
        \hline
         \multirow{2}{*}{Number of Python files} & Total & 17182 & 2983 & 2152 \bigstrut\\
          & Average & 59.87 & 80.62 & 58.16 \bigstrut\\
        \hlineB{2}
    \end{tabular}
\end{table}

To evaluate \tool, we used the Atcoder dataset \cite{nakamura2021atcoder}, which consists of 18644 Java solutions and 22317 Python solutions corresponding to 361 programming contest problems. To prove our concept, we created the train, validation, and test splits by dividing the dataset across different problems -- \ie train, valid, and test splits do not share the solution from the same problem. Since the problems are independent of each other, dividing in such a way prohibits data leakage across splits. Across all the experiments presented in the paper, we evaluate based on the same splits. \Cref{tab:data_stat} shows detailed statistics of the dataset. All experiments were performed with an A100 NVIDIA with 216 GB of RAM with a 2.44 GHz AMD 64-core processor.  

\subsection{Evaluation Metrics}

 we used three different evaluation metrics. The first is precision at N (PR@N), where N is between one and five inclusively. Precision at N is calculated as the average number of truly positive examples in the first N search results using every sample in the test set as a query against the rest of the test set. These metrics are benchmark standards in code search \cite{bishop2006pattern}, but we feel they are most important as they are indicative of the experience a user would have looking at a list of search results following a query. We also examined a metric termed Average Rank Gap (ARG) \cite{moffat2008average}, which is defined as the average rank of negative examples minus the average rank of positive examples. Unlike PR@N metrics, the ARG gives a sense of the entire set of search results available in the corpus. Between the PR@N metric and the ARG metric we get a sense of what a user would experience as well as \tool's performance across the entire dataset. 
 % \saikat{is this definition of AFP right?} 
 Lastly we evaluated the tools based on the average first position (AFP) \cite{willett1994some} of a positive result (lower is better). ARG provides context for the entire result set while AFP provides context for the highest ranked portion of the result set as a whole.

\section{Empirical Results}
\label{sec:results}
We answer the questions in the previous section one by one here. A full table of our results has been omitted for space consideration and clarity but is included in the supplementary material.

% \begin{figure*}
%     \centering
%     \includegraphics[width=\textwidth]{images/java-python.drawio.pdf}
%     \caption{Java to Python Search}
%     \label{fig:java-to-python}
% \end{figure*}

\subsection{RQ1 - Overall Performance}
\begin{figure}
  \centering
    \includegraphics[width=.9\linewidth]{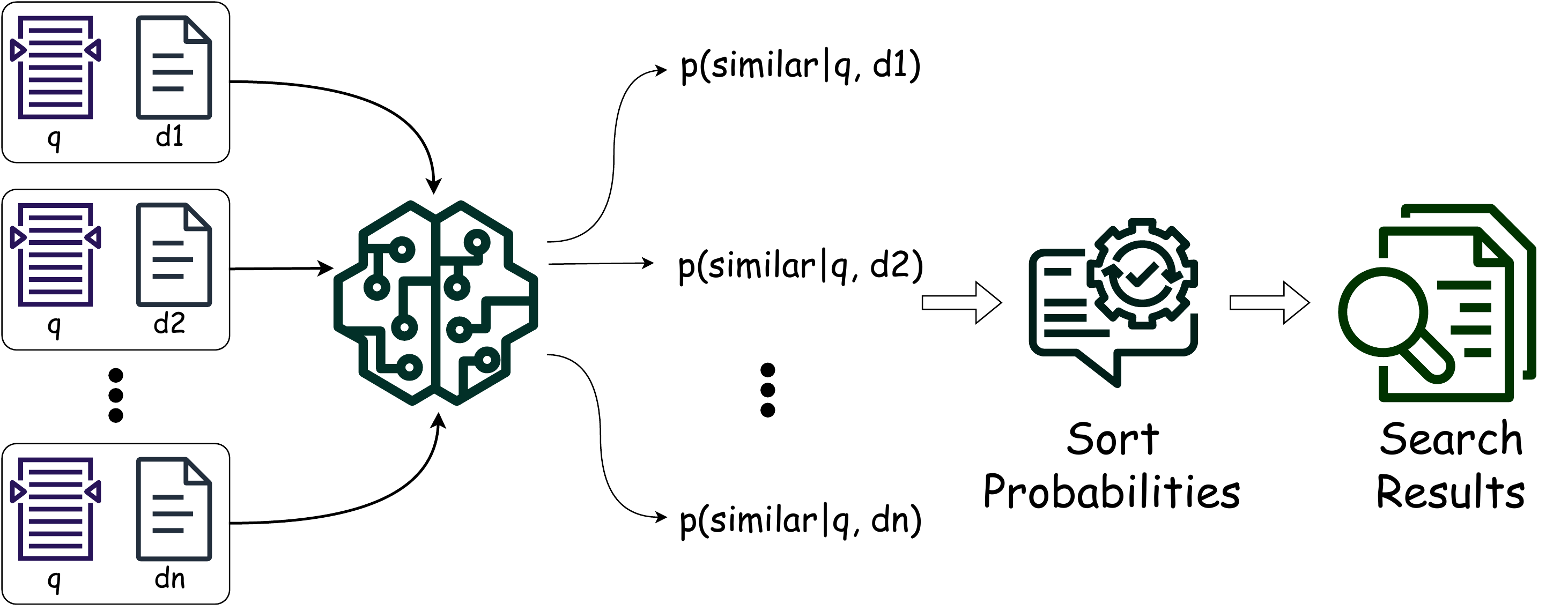}
    \caption{Classifier Style Search}
  \label{fig:traditional-search-methodology}
\end{figure}

The first thing we examined was the overall performance of the model, i.e., how good is the model at finding cross-language clones. Using the evaluation metrics defined in the previous section, we evaluated against both techniques that require training as well as those that do not. For strawman comparison we implemented a TF-IDF search with the BM25 python library \cite{macavaney2019bm25} that treats source code like natural language and performs a token-based search. We also implemented an AST-based search adapted from SLACC \cite{mathew2020slacc}, in which we take an AST representation of both the query and the document in the corpus, and convert them from language-specific AST's to a generic AST with common node types where possible. Then we use the standard tree distance algorithm \cite{zhang1989simple} to determine the similarity of the two samples. In the last of our non-training comparisons, we reimplemented COSAL's \cite{mathew2021cosal} token subset technique in which functions are divided into smaller snippets and the smaller snippets are clustered to determine a similarity between functions. We reimplemented this system because it is the state-of-the-art in cross-language code search, and we have different train, validate, test splits on the Atcoder dataset compared to their Atcoder experiments and we have a different experimental setup.

To compare against other code search techniques that require training, we selected three state-of-the-art LLM models that have shown promising results in a variety of code related tasks. CodeBERT \cite{feng2020codebert} is a pre-trained model for code representation and generation developed by Microsoft Research Asia. It can be fine-tuned for various downstream tasks and has achieved state-of-the-art performance on several benchmark datasets. CodeBERT's ability to transfer knowledge across programming languages makes it particularly useful for code generation tasks and has many potential applications in software engineering and programming language research.  GraphCodeBERT \cite{feng2020codebert} is an extension of CodeBERT that incorporates control flow graphs (CFGs) into its pre-training process. By using CFGs, GraphCodeBERT can better capture the relationships between code tokens and produce more accurate representations for code analysis and generation tasks. It has achieved state-of-the-art performance on several code analysis benchmarks, demonstrating its effectiveness in modeling complex code structures. UnixCoder \cite{UnixCoder} is a pre-trained model for natural language to shell command translation, developed by Facebook AI. It is trained on a large corpus of command-line usage examples and can generate shell commands given a natural language input. UnixCoder has achieved state-of-the-art performance on several benchmarks, demonstrating its ability to understand and translate complex natural language commands into executable shell commands. 

We implemented a search algorithm with each of the three models as described in Figure \ref{fig:traditional-search-methodology}, similar to that of \cite{CodeXGLUE}. The models are trained as a classifier on the training set and then for every document in the test set the classifier determines the probability of the document matching the given query and sorts results based on the highest probability. As such at inference time this process requires classifying the entire test corpus and incurs tremendous computational overhead. \tool in contrast can use a precomputed corpus embedding as shown in Figure \ref{fig:inference} and only needs to create an embedding vector from the query without concerning the rest of the corpus. To get the search results, simply choose the precomputed embeddings that are closest to the query vector. Performance experiments for our best version of \tool took roughly 14 hours while the other trained model code search comparisons would have taken multiple days to complete. To deal with this we randomly sampled 200 queries from the test set and reported our findings for those 200 samples. With this stratified sampling the experiments took roughly the same amount of time as the experiments with \tool. 

\begin{figure}
  \centering
  \begin{subfigure}[b]{0.4\textwidth}
    \includegraphics[width=\textwidth]{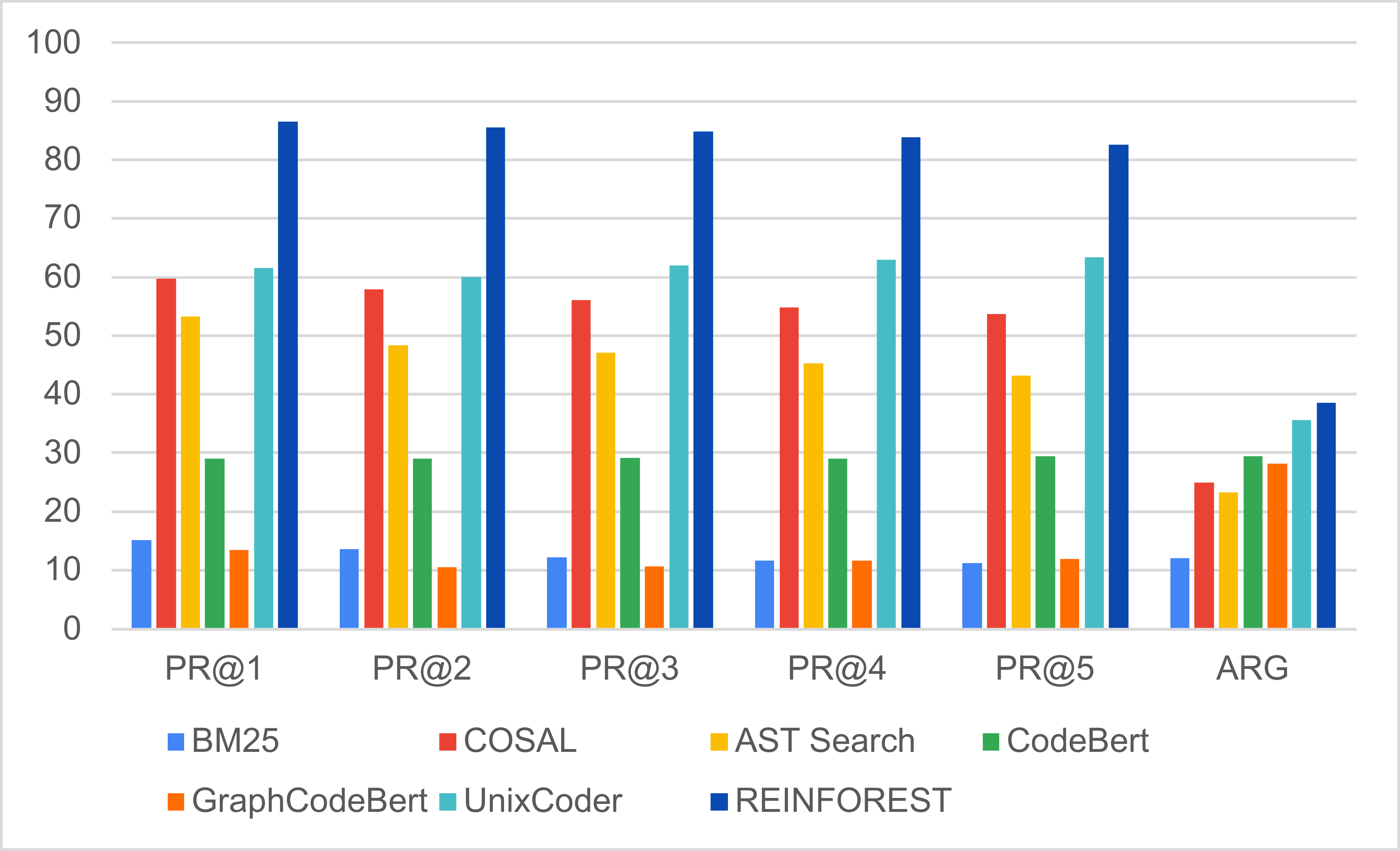}
    \caption{Java to Python Search}
    \label{fig:java_rq1}
  \end{subfigure}
  \hfill
  \begin{subfigure}[b]{0.4\textwidth}
    \includegraphics[width=\textwidth]{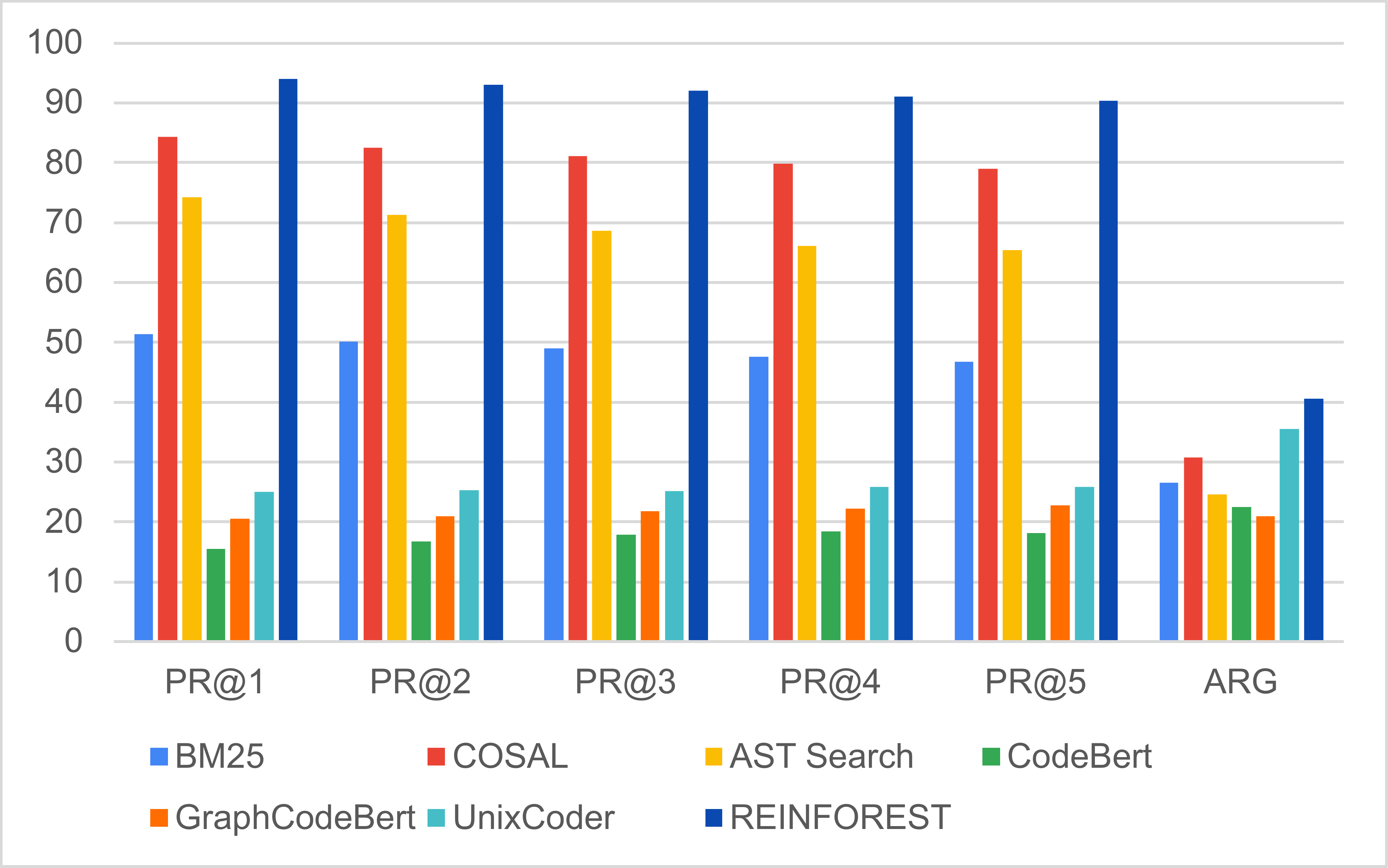}
    \caption{Python to Java Search}
    \label{fig:python_rq1}
  \end{subfigure}
  \caption{Overall \tool Performance}
  \label{fig:rq1-results}
\end{figure}

Figure \ref{fig:rq1-results} details the results of our experiments for PR@N metrics and ARG with the best-performing version of \tool against all the comparison techniques. For visual clarity, PR@N metrics were normalized to a percentage scale, ARG metrics were scaled by 100x, and AFP metrics were omitted but original values are included in the supplemental material. Figure \ref{fig:java_rq1} shows results for a Java query against a corpus of Python code while Figure \ref{fig:python_rq1} shows results for a Python query against a corpus of Java code. Across all evaluation metrics \tool outperforms all comparison techniques with improved performance up to 40\% compared to the next best performing method. Somewhat surprisingly we see the no training required techniques outperforming the trained techniques with the exception of UnixCoder and Java to Python search. While the relative performance of each method changes depending on the evaluation criteria, we find unsurprisingly that current state of the art COSAL generally is the next best performing.

Since \tool outperforms existing cross-language code search techniques by a considerable margin on all evaluation metrics we determine the answer to RQ1 as:

\RS{1}{\tool performs considerably better than existing techniques with a 40.6\% improvement over the nearest baseline for Java to Python search, and an 11.5\% improvement over the nearest baseline in Python to Java search. We outperform COSAL's state-of-the-art performance by up to 44.7\%.}

\subsection{RQ2 -- Model Generality}
While the results from the previous section show that the techniques described result in an effective tool, it remains unclear if the performance is because of the training techniques we employed or just the power of modern LLM's. To examine this we ask in RQ2 if our \tool techniques are effective in improving performance across many different LLM models both open-sourced and proprietary.  

For performance comparison we use the same metrics as described in the last subsection: PR@N, ARG, and AFP. We chose the most advanced available LLM models for comparison. First CodeBERT for the reasons described previously, but instead of training it as a classifier as we did in the first RQ, we fine-tuned its embedding on the cross-language code search task during the training procedure described in Section \ref{sec:method}. Additionally we ran our training models on OpenAI's Codex Ada, Babbage, Curie, and Davinci models. These are proprietary models and as such we could not perform fine-tuning during training, so we used the encodings from these models as is during training and \tool learns to maximize and minimize distances in these encoding spaces. Without access to the model internals the only discernible difference between the OpenAI models is the vector size of the embeddings, which are Ada: 1024, Babbage: 2048, Curie: 4096, and Davinci: 12188. As such we cannot explain why some OpenAI models may perform better than others.

To answer this question we ran two experiments. First we compared the performance of the various models' embeddings on cross-language code-to-code search without any alterations to \tool's performance on the same task training on the embeddings from those same models. Any increase in performance would have to be attributed to \tool's procedures instead of the base LLM. 

Second we compared the performance of \tool trained with the various underlying model embeddings on each evaluation metric to the next best performing code-to-code search technique. To understand the significance of the generality of our technique we should examine it relative to the state-of-the-art across all model architectures. Achieving state-of-the-art performance with all underlying models would further suggest that our techniques are valuable independent of the underlying LLM's.

\begin{figure}
  \centering
  \begin{subfigure}[b]{0.38\textwidth}
    \includegraphics[width=\textwidth]{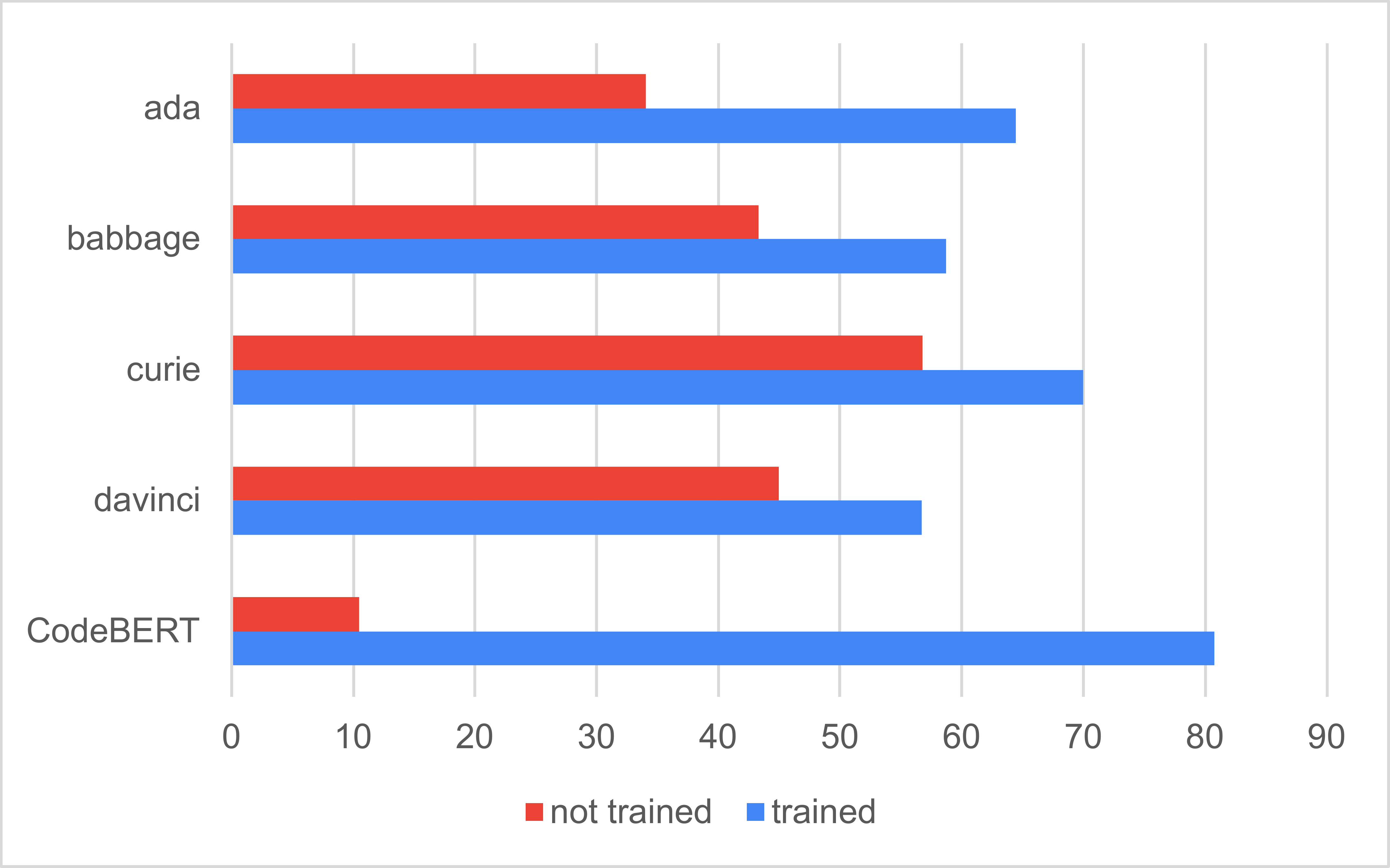}
    \caption{Java Training Improvement (PR@1)}
    \label{fig:rq2-java-training-impact}
  \end{subfigure}
  \hfill
  \begin{subfigure}[b]{0.38\textwidth}
    \includegraphics[width=\textwidth]{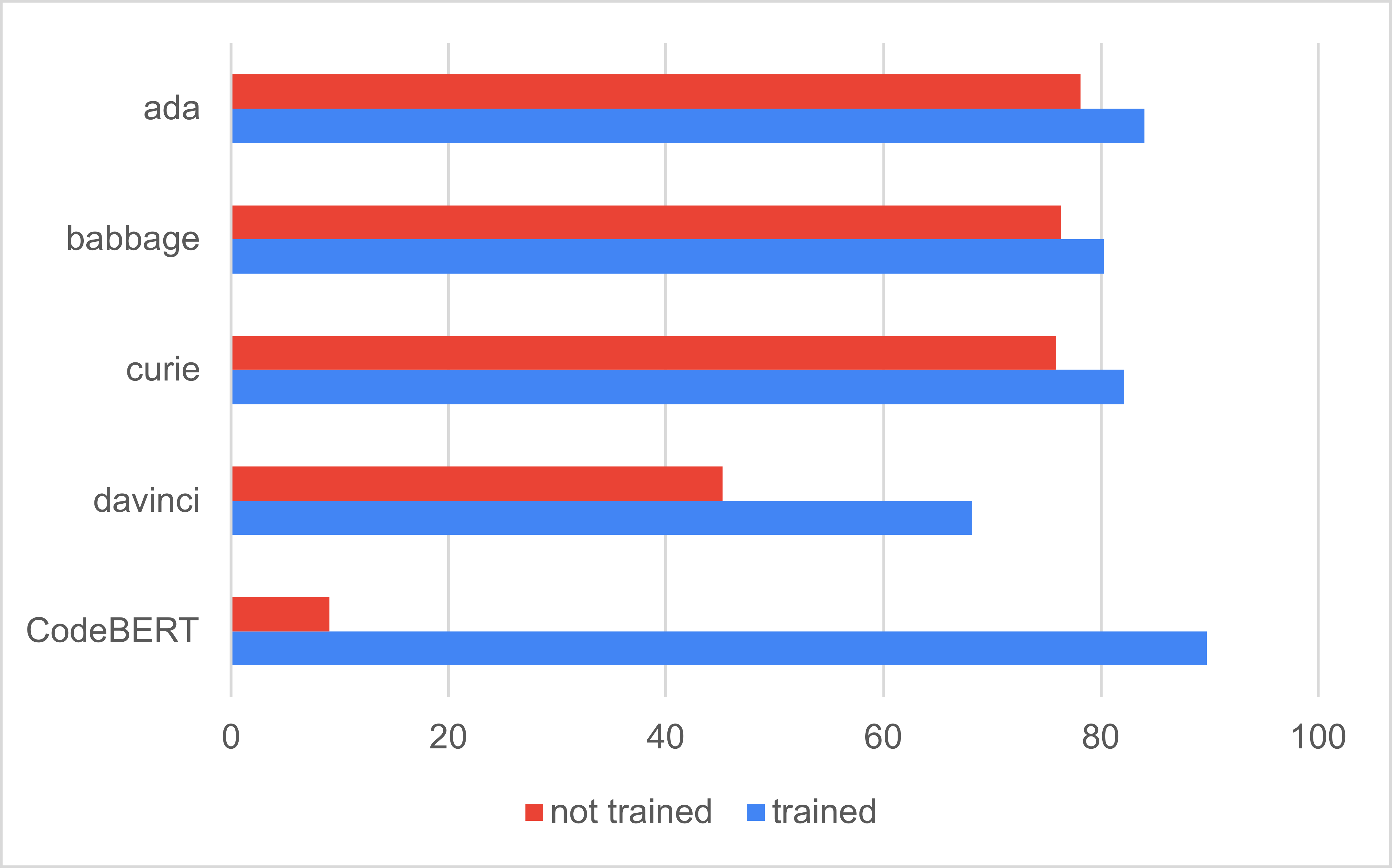}
    \caption{Python Training Improvement (PR@1)}
    \label{fig:rq2-python-training-impact}
  \end{subfigure}
  \caption{\tool Training Impact (PR@1)}
  \label{fig:rq2-training-impact}
\end{figure}

Figure \ref{fig:rq2-training-impact} compares performance between trained and untrained models. Clearly \tool's training always improves code-to-code search performance for both Java to Python and Python to Java queries. The improvements range from 5.2\% all the way up to 1044\%. Unsurprisingly our training has the largest impact on the one model that we could fine-tune during the training procedure. More interestingly, however, we find that despite having the weakest performance without training the fine-tuned CodeBERT model outperforms all other models on both Java and Python cross-language queries. This shows that fine-tuning the embedding during the training procedure has a larger impact than the size of the overall LLM and demonstrates the importance of fine-tuning. 

\begin{figure}
  \centering
  \begin{subfigure}[b]{0.41\textwidth}
    \includegraphics[width=\textwidth]{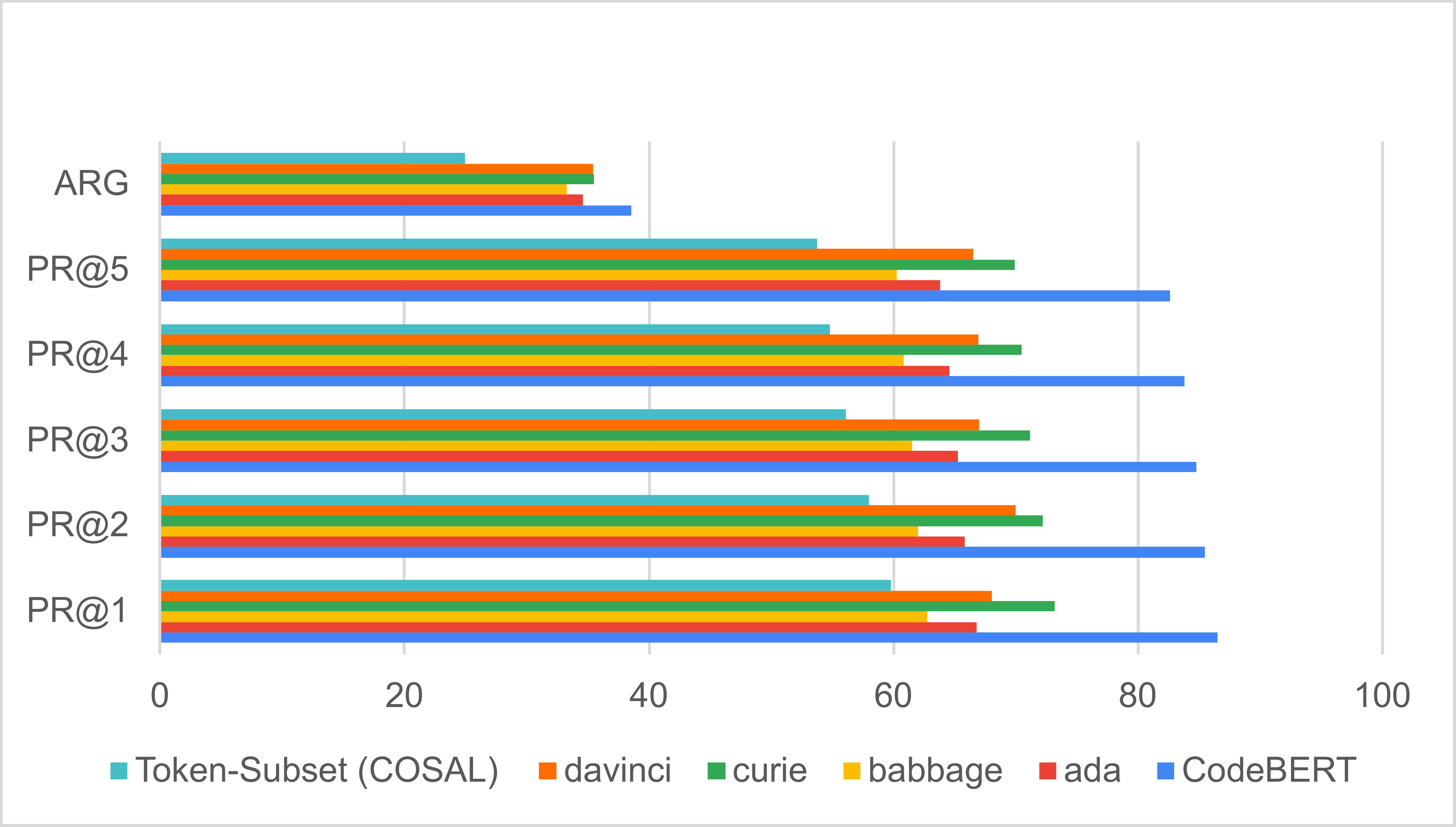}
    \caption{Java Model Performance}
    \label{fig:java_rq2}
  \end{subfigure}
  \hfill
  \begin{subfigure}[b]{0.41\textwidth}
    \includegraphics[width=\textwidth]{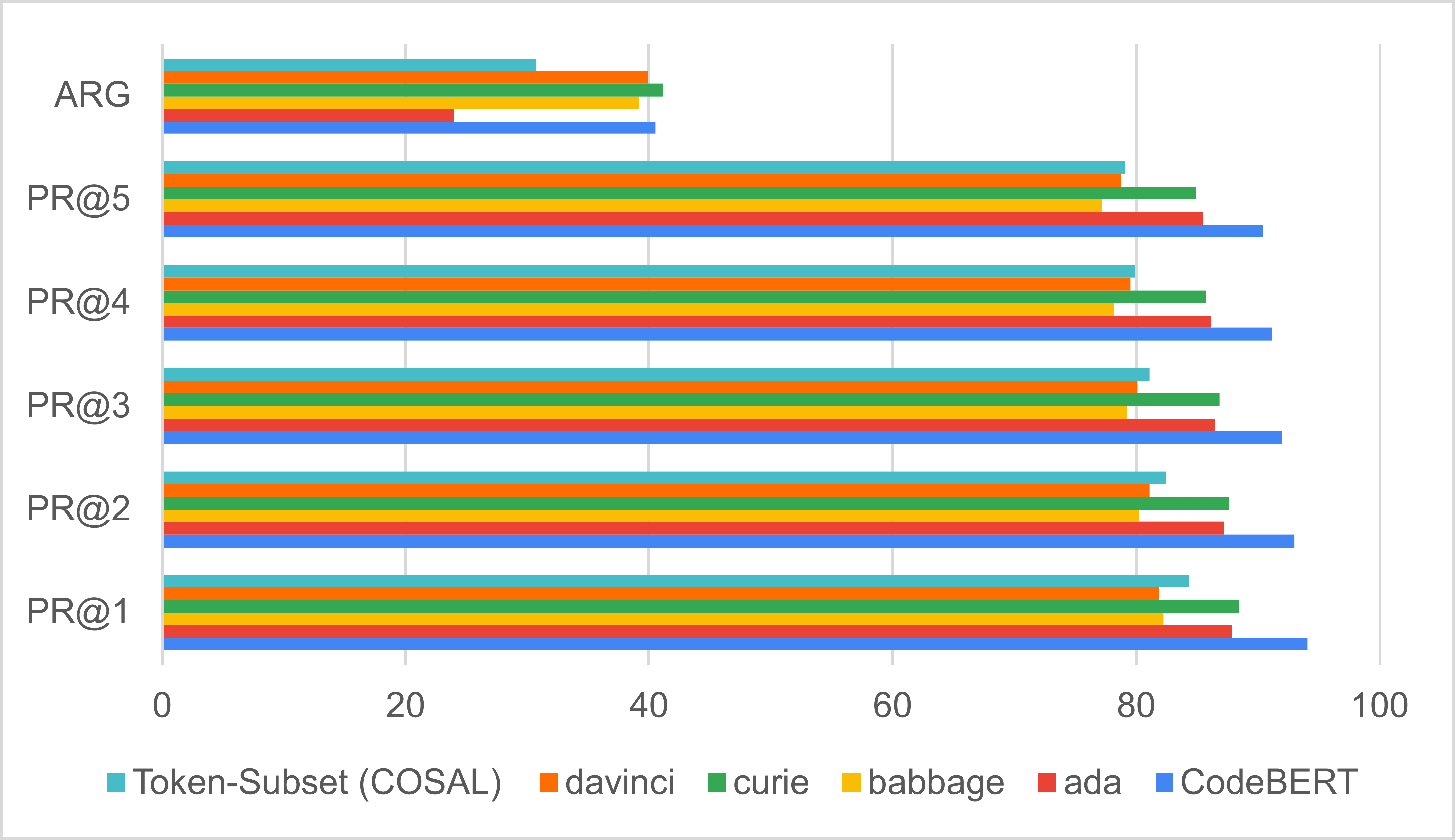}
    \caption{Python Model Performance}
    \label{fig:python_rq2}
  \end{subfigure}
  \caption{Performance of \tool across different models}
  \label{fig:rq2-performance-results}
\end{figure}

Figure \ref{fig:rq2-performance-results} shows the overall performance of the various models using \tool's techniques relative to COSAL (SOTA). Figure \ref{fig:java_rq2} details the results of Java to Python queries and shows that \tool achieves SoA performance across all metrics. Interestingly the same cannot be said for Python to Java queries as COSAL outperforms \tool in some circumstances but never out performs all versions of \tool. Since we do not know the internals of OpenAI's models it remains unclear why some of their models perform better than the SoA while others do not. However, most importantly we find that fine-tuned CodeBERT outperforms all other models on every metric for both cross-language queries. This solidifies fine-tuned CodeBERT as the highest performing version of \tool and demonstrates the power of open-sourced models. It also suggests that if the proprietary models were open-sourced we might see greater improvement.

The results from these experiments show 1) \tool's training techniques always improve the base LLM models and 2) \tool's techniques outperform the state-of-the-art in most circumstances and always when fine-tuning is possible. Thus we find that:

\RS{2} {\tool's techniques generalize across multiple LLM models, improving performance by a minimum of 26.2\% on Java to Python search and 5.17\% on Python to Java search, and when the base LLM is fine-tuned during training performance improves by 7.69x on Java to Python search and 9.96x on Python to Java search.}

\subsection{RQ3 -- Impact of SSS}
While the results from the previous RQ show that \tool's techniques are in fact responsible for the performance we saw in RQ1, we don't know how much of this performance is caused by the inclusion of runtime data in the form of the semantic similarity score (SSS). The SSS is the scored input output comparison described in Section \ref{sec:method}. 

To answer this question we compare the performance of trained versions of \tool when the SSS is included in the training procedure and when it is not. To perform these experiments we simply withheld the SSS from the training procedure. Any performance improvement would have to be from the model learning runtime behavior from the SSS. We tested this with CodeBERT as well as the OpenAI models to see if learning runtime behavior is in fact an inherent property of modern LLM's or peculiar to a single model.

\begin{figure}
  \centering
  \begin{subfigure}[b]{0.35\textwidth}
    \includegraphics[width=\textwidth]{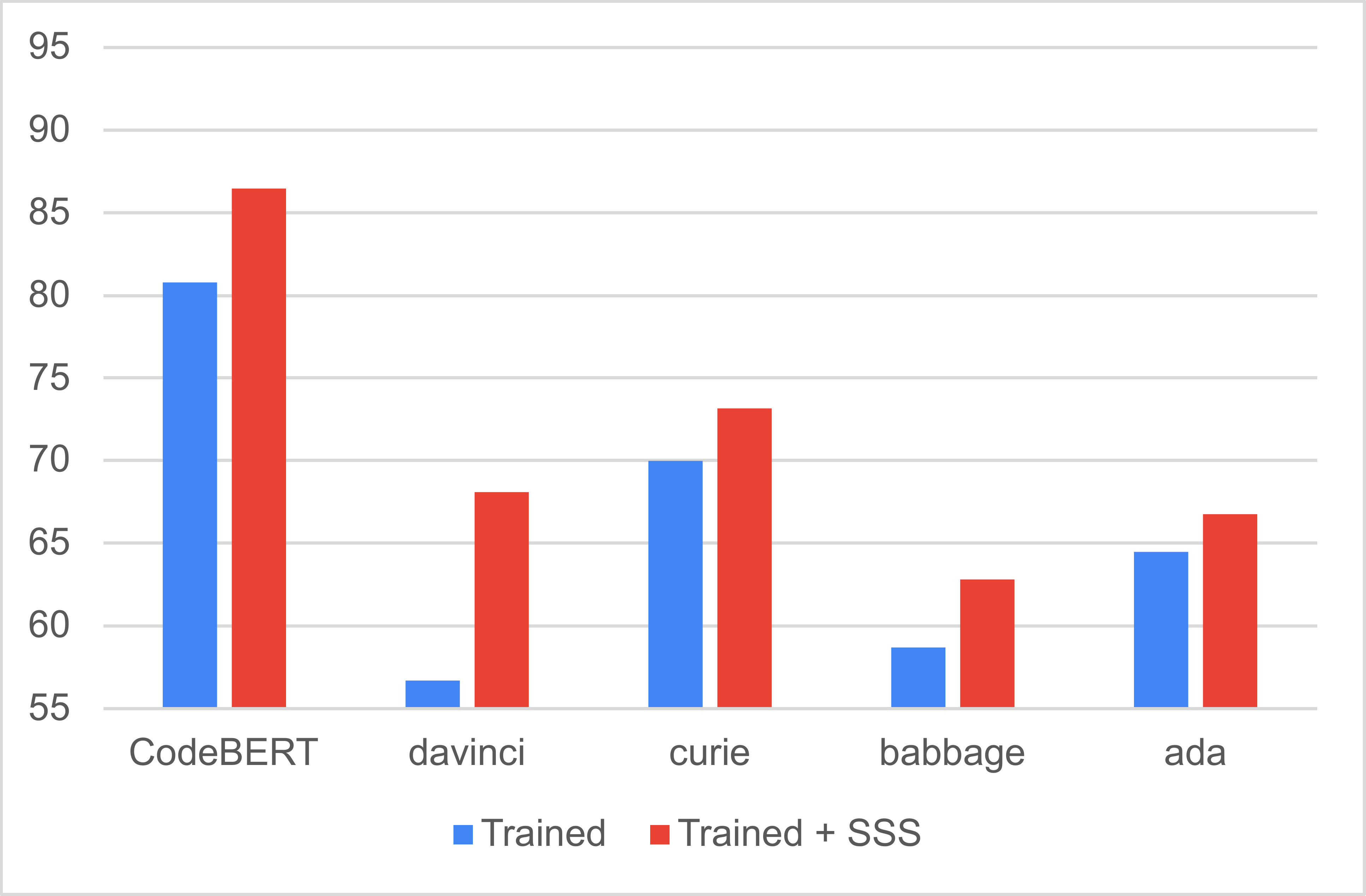}
    \caption{Java Impact of SSS (PR@1)}
    \label{fig:java_rq3}
  \end{subfigure}
  \hfill
  \begin{subfigure}[b]{0.35\textwidth}
    \includegraphics[width=\textwidth]{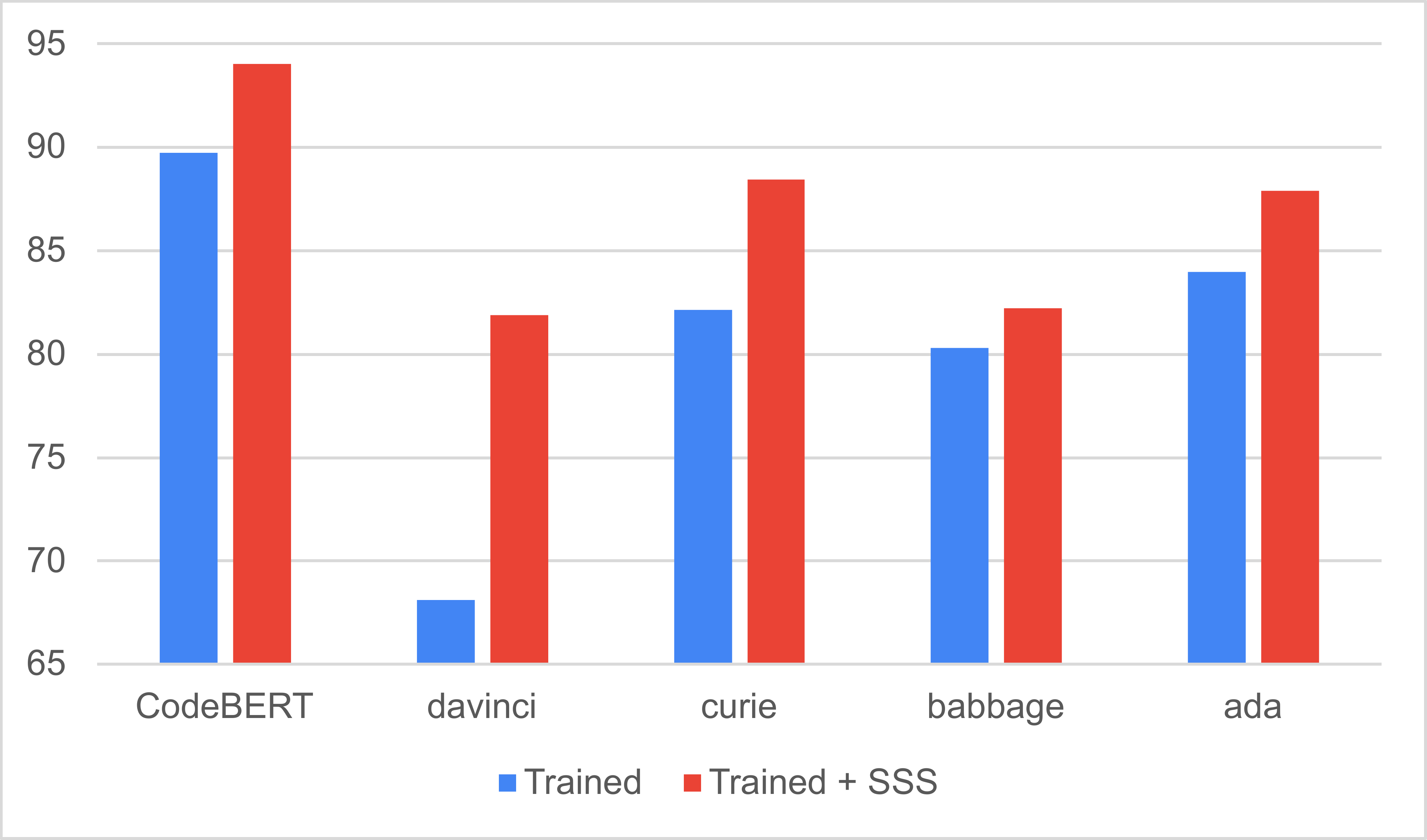}
    \caption{Python Impact of SSS (PR@1)}
    \label{fig:python_rq3}
  \end{subfigure}
  \caption{\tool SSS Impact (PR@1)}
  \label{fig:rq3-results}
\end{figure}

Figure \ref{fig:rq3-results} shows the impact of SSS on \tool's performance with various underlying models and cross-language queries. We find that including the SSS, regardless of the query language or the model used for embedding, always improves performance. This means that the models are in fact learning dynamic behavior despite never running the code during inference. Further it shows that this capability is inherent to modern LLM's and is not a property of a single model. We also see that it has the most improvement on one of OpenAI's proprietary models (davinci) which means that this technique continues to improve performance regardless of whether the model is fine-tuned during training or not. 

Thus we conclude that 
\RS{3} { Including SSS improves \tool's code search performance across all embeddings. Including the SSS with the best performing version of \tool (CodeBERT) contributed to a 7\% improvement for Java to Python queries and a 4.8\% improvement for Python to Java queries.}

\subsection{RQ4 -- Varying Training Samples}
Previous experiments show that 1) \tool's overall performance is better than the state-of-the-art, 2) is consistent regardless of the underlying model architecture, and 3) leverages runtime behavioral features. However it remains unclear if the model is learning from both positive and negative reference samples during training or if it is simply relying on one or the other. Further we have not yet investigated the impact of each reference sample on the model's performance. It seems plausible that increasing the size of the training set or adding more samples could improve performance. 

So we investigate RQ4 by varying the maximum number of positive and negative samples available during training. If the model is only learning from positive samples then performance should not dip when negative samples are removed and vice versa. Further if more training data would improve performance then if the number of reference samples available during training increases performance should increase as well. We ran performance experiments with the same number of positive and negative reference samples for one, three, and five references. Then we ran the experiments with zero positive references and five negative references and vice versa. we did this with a CodeBERT model fine-tuned during training and all available OpenAI Codex models off the shelf as in RQ2.   

\begin{figure}[t]
  \centering
  \begin{subfigure}[t]{0.45\textwidth}
    \centering
    \includegraphics[width=0.75\textwidth]{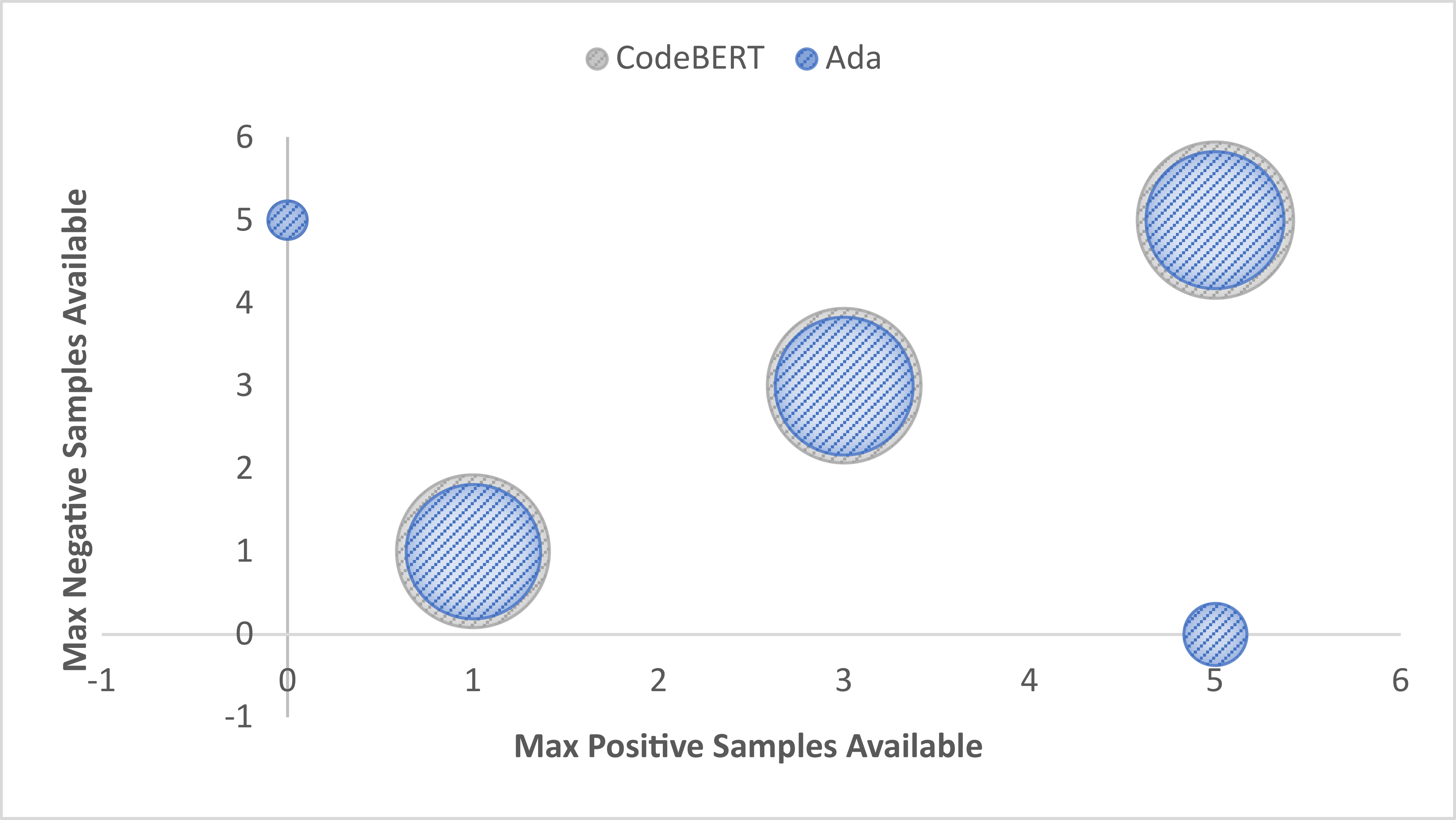}
    \caption{Java to Python Search varying reference samples (PR@1)}
    \label{fig:rq4-java}
  \end{subfigure}
  \hfill
  \begin{subfigure}[t]{0.45\textwidth}
  \centering
    \includegraphics[width=0.75\textwidth]{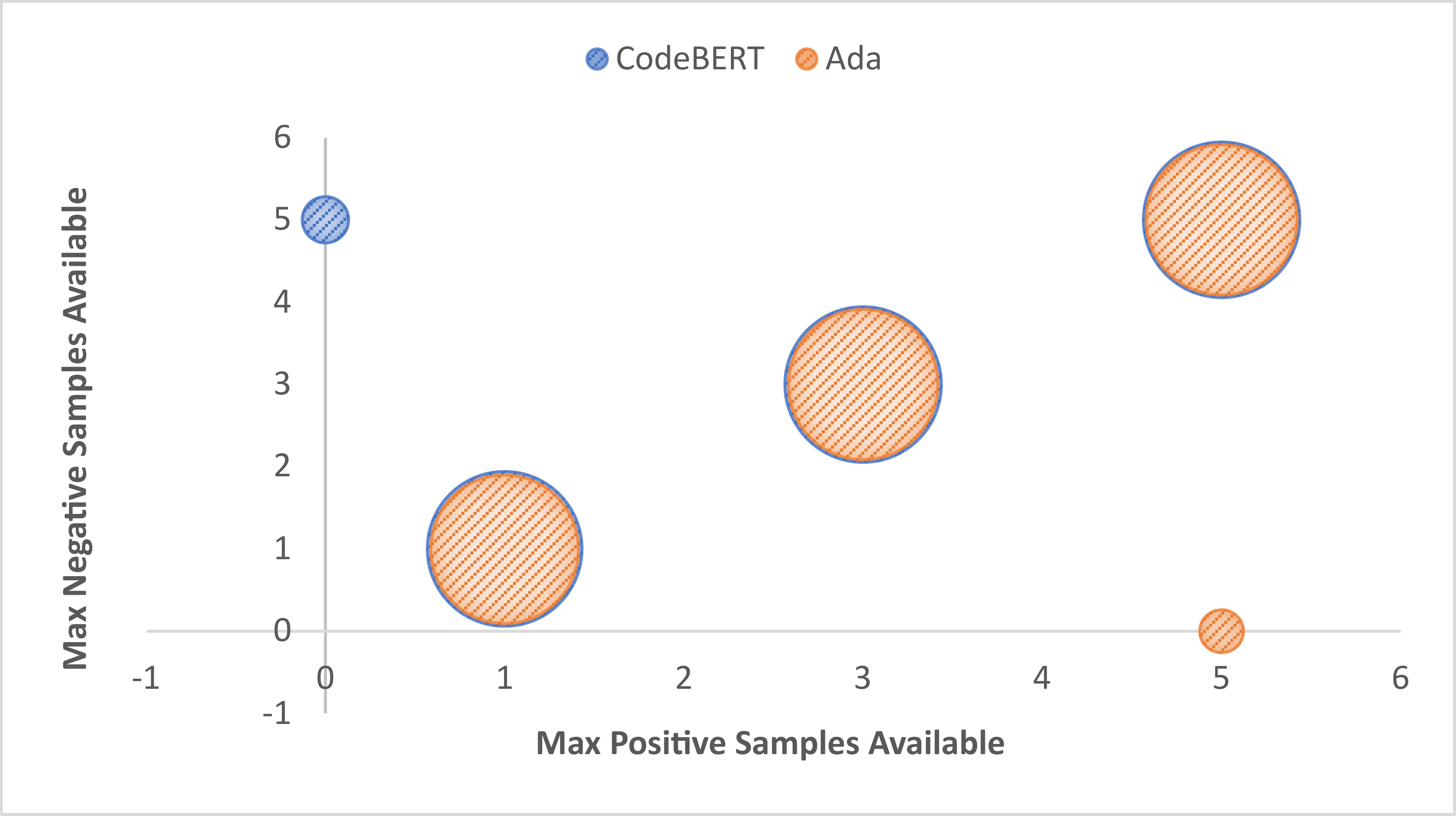}
    \caption{Python to Java search varying reference samples (PR@1)}
    \label{fig:rq4-python}
  \end{subfigure}
  \caption{\tool performance with varying number of reference samples during training}
  \label{fig:rq4-varying-samples}
\end{figure}

Full results of these experiments are included in the supplementary materials with all models and evaluation metrics. Figure \ref{fig:rq4-varying-samples} shows these results with both Java to Python search and Python to Java search for CodeBERT and OpenAI's Ada model evaluated on the PR@1 metric. The x and y axes show the number of positive and negative samples available during training and the size of the circles represents the PR@1 performance of the model in the circumstances described by the x and y axis. Larger is better. 

There are two interesting findings consistent in both models that can be fine-tuned and those that can not. The first is that when either positive or negative samples are omitted during training, performance decreases drastically. This indicates that \tool's dual maximizing and minimizing technique during training is in fact responsible for its performance. If only the positive samples were responsible for performance, then removing the negative samples would have no impact. This is clearly not the case. Second, we see that performance doesn't vary greatly when a single positive and negative reference sample are available. This means that increasing the size of the training set would not have meaningful impact on results. Still we do see a slight increase indicating it has some impact. This could be explained by the fact that during training source code labels are binary, which means that additional positive and negative samples do not explicitly provide extra context. With a different training metric that quantified the distance between samples more explicitly it is possible that the number of samples available would have more impact. However, this finding indicates that even in the case of a sparse dataset available during training, \tool's performance is still effective even if only a single positive and negative reference sample is available.      

\RS{4} {\tool's performance depends on both positive and negative samples. Using a combination of positive and negative samples improves performance by up to 10.2x and 17.8x for Python to Java search and 15.5x and 12.2x for Java to Python search, compared to only positive and negative references respectively. The addition of more than one positive and negative sample during training performance improvement decays, with a 5.4\% and 0.6\% improvement for Java to Python and Python to Java search, respectively.}

\section{Related Work}
\label{sec:related}

\subsection{Code Search}
Code search has been studied for a long time but has become increasingly more important with the rise of machine learning and the availability of large open-source code repositories like Github \cite{github}. Di Grazia et al. provide a comprehensive overview \cite{10.1145/3565971}. Techniques have generally fallen into two groups. The first relies primarily on static techniques that do not run the code either at training or inference time and the second also includes dynamic information for more context but takes on the additional overhead of needing to run code both in training and at inference time. Early static code similarity solutions like CCFinder \cite{1019480} and CP-Miner\cite{1610609} rely on token-based intermediate representations, and other early solutions like \cite{baxter1998clone} rely on graph-based and AST-based intermediate representations. Recently, FaCoy \cite{10.1145/3180155.3180187} uses code fragments to augment the code query to increase useful matches but encodes no behavioral information and requires previous queries to gain context. Pinku et al. show that additional information from a transcompiler can be powerful enough to take single language models and apply them across languages \cite{pinku2023pathways}. Adding such additional information remains cumbersome at inference time. Other tools focus on the ability to match structurally different but semantically similar code clones like CCGraph \cite{10.1145/3324884.3416541}, which leverages program dependency graphs and graph kernels to run machine learning directly on a graph representation of the code. These inherently expensive representations are difficult to train and do not encode any dynamic information. Another tool SourcererCC \cite{10.1145/2884781.2884877} focuses on such matches in large code bases, and while it scales well it fails to capture enough context to perform as well as other tools. Some security-focused code search tools like Macneto \cite{10.1145/3211346.3211352} apply clone detection techniques to an intermediate representation, e.g., examining Java byte code to detect clones after obfuscation. Such tools fall into a different category of clone detection outside of source code clone detection. 

The other set of code clone detection tools leverage dynamic behavior in their code representations. While this provides important semantic information, it adds complexity to the problem by requiring runnable code both during training and at inference time. With the modern intricacies of build environments this is not practical at scale.  Some recent work on dynamic code clone detection has moved towards taking code and splitting it into smaller segments and comparing the segments. This technique attempts to mitigate the difficulties of trying to match large code segments that are likely to behave similarly but have different structures, like COSAL \cite{mathew2021cosal} and SLACC \cite{9283951}. The technique is especially valuable in comparing code across languages. But splitting the code struggles to scale as the corpus grows larger, especially in a dynamic context that executes fragments repeatedly for behavioral information. HitoshiIO \cite{7503720} examines function input and output pairs to identify similar code and DyCLink \cite{10.1145/2950290.2950321} combines this information with code graph analysis and execution path information for a more informed but heavyweight solution.  To date, as far as we know, \tool is the first work that utilizes both static and dynamic information but does not incur the overhead of running code at inference time.

\subsection{Encoding Dynamic Information}
Encoding dynamic behavior in the model has been studied extensively \cite{he2016deeplog, allamanis2018learning, watter2015action, pham2015modelling}. The most similar work to ours that we know of is Trex \cite{trex}, which identifies semantically similar functions by encoding micro-execution traces during training and then transfers learning those values onto a pre-trained model. Despite the similar training/inference process there are several important differences between \tool and Trex: 1) Trex learns from binary level instruction sets, 2) during training it only maximizes agreement between similar sets and does not maximize distance between dissimilar samples, and 3) it would not work with proprietary models where transfer learning is not possible.
GraphCodeBERT \cite{guo2020graphcodebert} is another prior work that we compare to directly with the experiments in Section \ref{sec:results}. Execution specific dataflow is a dyanmic feature, but GraphCodeBERT extracts dataflow from the control flow graph without executing the code. Other similar works include Boost \cite{ding2022contrastive}, a contrastive learning approach that generates semantically identical code and trains on the generated examples. This work identifies behaviorally similar code but does so without considering runtime information.

Learning from dynamic behavior is especially prevalent in the security community as malware often obfuscates source code to avoid detection. Most learning approaches that leverage runtime information function like EnDroid \cite{8374030}, a malware detection agent for android that learns from dynamic features like execution traces and system activity, but requires collecting these dynamic features at inference time. Approaches like this take a considerably longer time to perform inference with since each sample must be examined relative to the query and runtime information must be collected for all queries. Another system, \cite{8681127}, suffers the same limitations.

\section{Threats to Validity and Limitations}
\label{sec:threats}
\subsection{Internal}
Since we had unique training, validation, and testing splits forming a singular experimental setting compared to results reported by the benchmark comparison techniques including COSAL, we were forced to implement the benchmark techniques ourselves as best we could. While the NLP-based search and the AST diff algorithm are fairly straightforward, recreating COSAL's snippet execution technique is somewhat complex. It is possible that some errors were made in recreating their results but the results in our evaluation are consistent with their findings. We evaluated benchmark techniques that involved training based on a stratified sampling of 200 randomly selected queries from the test set. It is possible that performance would differ when running the trained code search benchmarks over the entire dataset.

We evaluated on the Atcoder dataset in order to compare with COSAL's state of the art performance, and in order to compile the SSS as many samples as possible were executed but we were unable to execute much of the dataset since it did not come with input specifications. When an SSS could not be collected we used a default value. It is possible that SSS could have a bigger or smaller impact if we had been able to generate inputs for the entire dataset. Furthermore, during SSS collection there were some sample pairs that were labeled as true positives but had different input types since the dataset allowed identical solutions with equivalent but different input structures (i.e. set vs list). These behavioral similar pairs scored an SSS of 0 during training and finding a true SSS for these pairs may change results.

We found that setting $\alpha$ to 0.2 had the best results of the values we tried, but more rigorous hyper-parameter tuning could result in a different optimal value.

\subsection{External}
The Atcoder dataset is compiled from submissions to a coding contest by practitioners of all levels. It remains unclear if the code in these contests is actually reflective of production code. Further we examine code-to-code search when two functions are defined to solve the exact same problem. It is unlikely to find such examples in practice and it remains unclear how well \tool would perform in real-world scenarios. As was mentioned earlier \ref{sec:method} we only consider behavioral clone, or not, during training and this limits the amount of context the model can learn from multiple positive and negative samples. Our experiments show the presence of both positive and negative samples impact performance but with a different training procedure that considered the distance between samples rather than classifying them as one of two categories the addition of more available samples during training may have more impact on performance. 

\subsection{Construct}
We chose our evaluation metrics based on what we believed to be standard and meaningful. There are other important evaluation metrics like recall, overall accuracy, and F1 score that we did not measure. We chose cross-language search because we believe that it was the most meaningful problem in identifying behavioral similarity, but experiments on same-language search would need to be carried out explicitly to evaluate \tool's effectiveness for that purpose.

\subsection{Limitations}
Finding appropriate inputs that can be used to calculate the SSS between two pairs of functions is analagous to input (test case) generation which remains an open research problem. Further our method depends on a single function to function mapping which may not occur in practice, as the same behavior can be divided between multiple functions in different codebases.

\section{Conclusion}
\label{sec:conclusion}
This paper presents a novel code-to-code search technique and an implementation of that technique called \tool. The technique leverages modern LLM's and enhances them by encoding runtime behavior in the form of a Semantic Similarity Score (SSS). Unlike other code search techniques that only consider positive samples during training, \tool both minimizes the distance between similar samples and maximizes the distance between dissimilar samples. We evaluated \tool on cross-language code search, both searching a corpus of Python samples with Java queries and vice versa. We found that our technique outperforms the state of the art on all evaluation metrics. Further, we found that our training technique always improves performance over modern LLM encodings out of the box, regardless of the underlying model, and we found that \tool performs especially well when the LLM models can be fine-tuned. This proves that our procedures presented here are the source of the improved performance and not the underlying models. We also present experiments quantifying both the impact of SSS and the importance of both positive and negative samples available during training. We found that including the SSS always improves performance, indicating the model does in fact learn dynamic behavior without dynamic context available at inference time, and that performance depends on the presence of both positive and negative samples. The complete set of results including results that were omitted for clarity is included in the supplementary material and all models and training procedures will be open-sourced.      

% \newpage

% \input{body/99.help_and_todo}
% \section*{Acknowledgement}

\bibliographystyle{IEEEtran}
\bibliography{main}
\balance

% \newpage

% \appendix
% \input{appendadix/main}
% \newpage
% \input{todo}
\end{document}